\documentclass[manuscript]{aastex}

\newcommand{\Kepler}{{\it Kepler~}}

\shorttitle{\Kepler Eclipsing Binary Stars}
\shortauthors{Pr\v sa et al.}

\begin{document}

\title{\Kepler Eclipsing Binary Stars. I. Catalog and Principal Characterization of 1879 Eclipsing Binaries in the First Data Release}

\author{Andrej Pr\v sa}
\affil{Villanova University, Dept.~of Astronomy and Astrophysics, 800 E Lancaster Ave, Villanova, PA 19085}
\email{andrej.prsa@villanova.edu}

\and

\author{Natalie Batalha}
\affil{San Jos\'e State University, One Washington Square, San Jos\'e, CA 95192}

\and

\author{Robert W.~Slawson and Laurance R.~Doyle}
\affil{SETI Institute, 189 N Bernardo Ave, Mountain View, CA 94043}

\and

\author{William F.~Welsh and Jerome A.~Orosz}
\affil{San Diego State University, 5500 Campanile Dr., San Diego, CA
92182-1221 USA}

\and

\author{Sara Seager}
\affil{Massachusetts Institute of Technology, 77 Massachusetts Ave, Cambridge, MA 02139}

\and

\author{Michael Rucker and Kimberly Mjaseth}
\affil{San Jos\'e State University, One Washington Square, San Jos\'e, CA 95192}

\and

\author{Scott G.~Engle and Kyle Conroy}
\affil{Villanova University, Dept.~of Astronomy and Astrophysics, 800 E Lancaster Ave, Villanova, PA 19085}

\and

\author{Jon Jenkins and Douglas Caldwell}
\affil{SETI Institute/NASA Ames Research Center, Moffett Field, CA 94035, USA}

\and

\author{David Koch and William Borucki}
\affil{NASA Ames Research Center, Moffett Field, CA 94035, USA}

\begin{abstract}
The \Kepler space mission is devoted to finding Earth-size planets orbiting other stars in their habitable zones. Its large, 105 square degree field-of-view features over 156,000 stars that are observed continuously to detect and characterize planet transits. Yet this high-precision instrument holds great promise for other types of objects as well. Here we present a comprehensive catalog of eclipsing binary stars observed by \Kepler in the first 44 days of operation, the data being publicly available through MAST as of 6/15/2010. The catalog contains 1879 unique objects. For each object we provide its \Kepler ID (KID), ephemeris (BJD${}_0$, $P_0$), morphology type, physical parameters ($T_\mathrm{eff}$, $\log g$, $E(B-V)$), the estimate on third light contamination (crowding), and principal parameters ($T_2/T_1$, $q$, fillout factor and $\sin i$ for overcontacts, and $T_2/T_1$, $(R_1+R_2)/a$, $e \sin \omega$, $e \cos \omega$, and $\sin i$ for detached binaries). We present statistics based on the determined periods and measure the average occurence rate of eclipsing binaries to be $\sim$1.2\% across the \Kepler field. We further discuss the distribution of binaries as a function of galactic latitude, and thoroughly explain the application of artificial intelligence to obtain principal parameters in a matter of seconds for the whole sample. The catalog was envisioned to serve as a bridge between the now public \Kepler data and the scientific community interested in eclipsing binary stars.
\end{abstract}


\section{Introduction}

The contribution of binary stars and, in particular, eclipsing binaries (hereafter EBs) to astrophysics cannot be overstated.  EBs can provide fundamental mass and radius measurements for the component stars (e.g.\ see the extensive review by \citealt{andersen1991}).  These mass and radius measurements in turn allow for accurate tests of stellar evolution models (e.g.\ \citealt{pols1997}; \citealt{schroder1997}; \citealt{guinan2000}; \citealt{torres2002}).  In cases where high quality radial velocity curves exist for both stars in an EB, the luminosities computed from the absolute radii and effective temperatures can lead to a distance determination.  Indeed, EBs are becoming widely used to determine distances to the Magellanic Clouds, M31, and M33 \citep{paczynski1997, paczynski2000, guinan1998, fitzpatrick2002, wyithe2001, wyithe2002, bonanos2003, bonanos2006, north2010}.

Large samples of EBs have been generated as byproducts of automated surveys for microlensing events (e.g.\ OGLE, \citealt{udalski1998, wyrzykowski2004, udalski2004}; MACHO, \citealt{alcock1997}; EROS, \citealt{grison1995}) and automated searches for gamma-ray burst afterglows \citep{akerlof2000}.  In addition, the {\em Hipparcos} mission provided a sample of binaries (both astrometric and eclipsing; \citealt{perryman1997}). Other ground-based projects yielded databases of EBs ready for data mining \citep{devor2008,christiansen2008}. Large samples are useful to determine statistical properties and for finding rare binaries (for example binaries with very low mass stars, binaries with stars in short-lived stages of evolution, very eccentric binaries that show large apsidal motion, etc.).  The catalogs of EBs from the ground-based surveys suffer from various observational biases such as limited accuracy per individual measurement and complex ``window'' functions (e.g.\ observations can only be done during nights with clear skies and during certain seasons).  The {\em Hipparcos} mission had all-sky coverage with good photometric precision, but had limited sampling for stars brighter than magnitude 8.

The {\Kepler} mission will provide essentially continuous coverage of $\sim 156,000$ stars with unprecedented photometric precision.  We present here a catalog of EBs in the first data release (``Q0'' and ``Q1'').  This catalog will serve as a bridge between the public \Kepler data and members of the wider scientific community interested in eclipsing binary stars.

\section{Observations}

The details of the \Kepler spacecraft and photometer have been presented elsewhere, e.g. \citet{borucki2010a, koch2010, batalha2010, caldwell2010, gilliland2010, jenkins2010b, jenkins2010a}, and others, but for completeness we give a synopsis here.

The \Kepler spacecraft is in a heliocentric, Earth-trailing orbit, allowing for near continuous coverage of its 105 square-degree field of view (FOV). The telescope is a 0.95-m Schmidt-design with a 1.4-m f/1 primary, designed to allow \Kepler to monitor $\sim$156,000 stars of interest simultaneously and continuously for the duration of the mission. The hardware design stresses simplicity to minimize risk: there are no moving parts other than the dust cover (ejected during spacecraft commissioning), primary mirror focus, and gyro reaction wheels -- there is no filter wheel and no shutter. The lack of shutter means starlight continues to illuminate the CCD during readout, but the effect of the smearing is measured via 20 rows of masked pixels and is removed in the data calibration pipeline.

The photometer camera contains 42 CCDs with $2200 \times 1024$ pixels, where each pixel covers 4~arcsec. The \Kepler point spread function has an average 95\% encircled energy width of 4 pixels diameter, and for a star centered on a pixel an average of 47\% of the flux is recorded in the central pixel \citep{bryson2010}. Because \Kepler is telemetry-limited, not every pixel is read out and stored; only pre-selected stars of interest are observed, with 32 pixels per star recorded on average; the optimal aperture photometry uses approximately half of the recorded pixels. The \Kepler FOV, centered at $RA=19^{h}22^{m}40{s}$ and $Dec=44^{o} 30' 00"$ in the Cygnus-Lyra region, spans galactic latitudes 5-22 deg N and is rich with stars. This region was chosen for a variety of reasons, one of which is that being slightly off the galactic plane greatly reduces the number of distant giants in the FOV. Of the roughly half million targets in the FOV brighter than 16th mag, approximately 30\% have been targeted. The selection criteria conformed to the primary goal of the mission: to measure the fraction of stars that have terrestrial planets near their habitable zones. Stars where such a signature is impossible to detect (i.e.~giants, stars fainter than 16th mag, stars in over-crowded fields) were omitted from the target list. All previously known eclipsing binaries in the FOV were included in the target list (383 targets; see \S 3.1). Of the 156,097 stars observed in Q1, $\sim$60\% are G-type stars on or close to the Main Sequence. Spectral classification and stellar parameters were estimated using dedicated pre-launch ground-based optical multi-color photometry plus 2MASS J, K, H magnitudes, matched to the \citet{castelli2004} model stellar atmospheres and \citet{girardi2000} evolutionary tracks. This information, along with information from the USNO-B, Tycho, and Hipparcos catalogs are presented in the {\it Kepler Input Catalog} (KIC --- see \citealt{batalha2010}), available at MAST\footnote{Multi-Mission Archive at Space Telescope Science  Institute; http://archive.stsci.edu/kepler}, and much of this information is also included in the fits file headers. Because these are based on photometry, not spectroscopy, the surface gravity ($\log g$) and metallicity [Fe/H] estimates can be quite uncertain, and a 25\% error in the quoted stellar radius is possible. The KIC contains $\sim$13.2 million targets, of which $\sim$4.4 million fall into the \Kepler FOV; however, not every \Kepler target will have a KIC designation.\footnote{The IAU designated naming convention is to use the \Kepler IDentification number, KID. The KID's follow the KIC numbers when possible.} For additional details of the target selection procedure (and information toward any biases that may be present in the EB catalog), see \citet{batalha2010}.

The CCDs are read out every 6.54~s (6.02~s live-time) and then summed into 29.4244~min Long Cadence bins. In addition, up to 512 targets can be observed in Short Cadence mode, at a 59-sec sampling. The bandpass spans 423--897 nm, chosen to limit the effect of the variable Ca~II H~\&~K lines and fringing in the infrared. Thus the \Kepler passband spectral response is similar to a broad V+R bandpass, and \Kepler magnitudes, $K_{p}$, are usually within 0.1 of the R-band magnitude. Simple aperture photometry is used to measure star fluxes, and the light curves are given in electrons per cadence length. The effective dynamic range is $K_{p}=7-17$ mag, though even fainter targets have reliable photometry \citep{gilliland2010}. Targets with $K_{p} < 11.3$ saturate the CCD in the 6.02~s exposures; however, this does not thwart the relative precision. The CCD clock voltages are set such that no $e^{-}$ is left behind, and because of the exceptional stability of the photometer platform, due in part to its Earth-trailing heliocentric orbit, precise photometry is possible well above the pixel saturation level. The \Kepler design goal was to achieve a photometric precision of 20~ppm for a 6.5 hour exposure of a G2-type $V=12$ target, and initial estimates of the instrument performance indicate that \Kepler is approaching that goal \citep{koch2010}.

Four times a year the \Kepler spacecraft rolls by 90 degrees to re-align its solar panels, and these define epochs known as ``Quarters''. The 9.7 days of data acquired during the end of spacecraft commissioning are known as ``Q0'' data, and the first operational dataset as ``Q1''. The duration of Q1 is shorter than the nominal quarter duration because the launch on 2009~Mar~6 necessitated a roll only 33.5-d after the start of Q1. Q0 and Q1 together span 43 days, from 2009 May 02--Jun 15 UT. The 52,496-star Q0 set is different from all subsequent sets because its primary intended use was to measure \Kepler performance and examine stars that were initially excluded from the \Kepler target list. All uncrowded stars with $K_{p} < 13.6$ mag, of all spectral types and luminosity classes, are included, with the exception of 160 stars brighter than $K_{p} = 8.4$ mag.

As with any instrument, there are artifacts and features unique to {\it Kepler}. We strongly encourage users of \Kepler data to read the  {\Kepler Instrument Handbook}, the \Kepler {\it Archive Manual} and the most current version of the {\it Data Release Notes} \citep{vancleve2009}, all of which are available at MAST. Of particular note, the data available at MAST contain ``RAW'' and corrected ``CORR'' aperture photometry. The raw observations do include pipeline processing, but not as much as the more heavily processed ``CORR'' data. Users of \Kepler data should obtain and work with both versions of the light curves and use whichever is better suited to their goals. Since the purpose of \Kepler is to find Earth-like planets, 
the calibration pipeline is optimized towards that goal. As consequence, the current version of the pipeline often over-filters the corrected fluxes in Q0, so these should be used with extreme caution for measuring anything other than periods and epochs. For these reasons, the ephemerides in this catalog are derived from Q0+Q1 data, but the principal parameters are based entirely on Q1 observations.

The detection of the EBs presented in this paper are based primarily on the initial pipeline calibrated light curves. The revised calibration (SOC Data Release \#5) coincides with the public release of the Q0 and Q1 data and the release of this catalog. While every attempt has been made to use the more recent calibration, the data release schedule did not permit full re-analysis of all aspects of all systems. Fortunately the differences in calibration are generally minor (with the exception of the corrected Q0 data noted above) and do not affect the results presented here. A most notable difference between the observations used in this work and the revised calibration are the times recorded in the fits headers: the initial calibration provided modified Julian times with the applied heliocentric correction (MJD), while the recent calibration provides times corrected for barycentric motion (BJD). We have converted the MJD to BJD using the interpolation formulae provided in the SOC Data Release \#4. We used the center of the \Kepler FOV to determine the linear correction and apply it to the provided MJD. The accumulated correction to MJD over the Q0/Q1 duration is $\sim 160$\,s, implying that the derived periods without this correction would be anomalous to $\sim 4 \cdot 10^{-5}$. The difference between the interpolated and the true BJD time is accurate to $\sim 10^{-7}$\,days, which is well below the accuracy of the determined period to cause any systematic effects. Since all post-Q1 data will be delivered with BJD time-stamps, further ephemeris refinement will not suffer from any interpolation artifacts.

In addition to the photometry contained in the .fits files, users are encouraged to examine the pixel row and column positions of the centroids of the starlight. This information can be valuable when considering dilution or contamination of the EB. For example, if there are two stars within the aperture, one of which is an EB and the other a source of background light, then during eclipse the ratio of light from the EB compared to the contaminating star decreases, and the center of light will shift a small fraction of a pixel toward the contaminating star. Changes in positions in a single Long Cadence observation as small as a 
millipixel and better are measureable \citep{monet2010}. Plots of flux versus pixel position (known as ``rain plots'') have been useful for rejecting false extrasolar planet candidates that were in fact diluted eclipsing binaries \citep{jenkins2010c}. Finally, Full-Frame Images of the 42 CCDs are also available. Although of low spatial resolution, the image can be examined in cases where it is suspected that light from a neighboring bright star may be leaking into the aperture of the target star, either diluting the signal, or inducing a spurious signal if that contaminating star is variable.

\section{EB Catalog}

There are two strong scientific cases for building the catalog of EBs. The first one is obvious: the unprecedented quality and uninterrupted sampling of \Kepler data are a leap forward in being able to perform modeling and analysis of those stars. The second is to estimate the occurence rate of EBs across the \Kepler field. This bears special significance for the \Kepler core mission of finding planets, since occurence rate provides rough estimates on the contamination statistics that translates into the false positive probability (see \citealt{borucki2010b}). In this section we describe the sources and design of the catalog.

\subsection{Detection}

The pre-launch \Kepler target list included 383 known EBs. 59 were found from the SIMBAD Astronomical Database using a query for morphological type within the \Kepler field of view (FOV). The All Sky Automated Survey - North \citep{pigulski2008} cataloged over 1000 variable stars within the \Kepler FOV of which 127 were added as target EBs. An additional 7 EBs were added from the Hungarian-made Automated Telescope variability survey of the \Kepler FOV \citep{hartman2004}. The remaining pre-launch targets were identified from an analysis of the survey 
conducted by Vulcan -- a 10cm aperture, wide-field, automated CCD photometer \citep{borucki2001}. Approximately 60,000 stars were observed in and around the \Kepler field of view for a period of 60-97 nights. Automated transit detection via matched-filter correlation \citep{jenkins1996} yielded some 600 eclipsing binary detections, 190 of which are in the \Kepler field and, consequently, added to \Kepler's pre-launch target list \citep{mjaseth2007}.

As part of the main processing pipeline, \Kepler data are passed through the Transit Planet Search (TPS) algorithm. The pipeline searches through each systematic error-corrected flux time series for periodic sequences of negative pulses corresponding to transit signatures. The approach is a wavelet-based, adaptive matched filter that characterizes the power spectral density (PSD) of the background process that yields the observed light curve and uses this time-variable PSD estimate to realize a pre-whitening filter and whiten the light curve \citep{jenkins2002,jenkins2010d}. TPS then convolves a transit waveform, whitened by the same pre-whitening filter as the data, with the whitened data to obtain a time series of single event statistics. These represent the likelihood that a transit of that duration is present at each time step. The single event statistics are combined into multiple event statistics by folding them at trial orbital periods ranging from 0.5 days to as long as one quarter ($\sim$90 days). The step sizes in period and epoch are chosen to control the minimum correlation coefficient between neighboring transit models used in the search so as to maintain a high sensitivity to transit sequences in the data. The transit durations used for TPS through June 2010 were 3, 6 and 12 hours. These transit durations are being augmented to include 1.5, 2.0, 2.5, 3.0, 4.0, 5.0, 6.0, 7.5, 9.5, 12.0, and 15.0 hours. TPS is also being modified to combine  to conduct searches across the entire mission duration by "stitching" quarterly segments together. Since eclipsing binaries often exhibit periodic pulse trains with durations from a few hours to half a day, most are identified by TPS.

The maximum multiple event statistic is collected for each star and those with maximum multiple event statistics greater than $7.1 \sigma$ are flagged as Threshold Crossing Events (TCEs). The Data Validation (DV) pipeline fits limb-darkened transit models to each TCE and performs a suite of diagnostic tests to build or break confidence in each TCE as a planetary signature as opposed to an eclipsing binary or noise fluctuation \citep{wu2010,tenenbaum2010}. DV removes the transit signature from the light curve and searches for additional transiting planets using a call to TPS. TCEs with transit depths more than 15\% are not processed by DV, as well as light curves whose maximum multiple event statistics are less than 1.25 times the maximum single event statistic.

The TPS output for Q0/Q1 data yielded $\sim$10,000 TCEs above the 7.1-$\sigma$ threshold that have been sifted for EB candidates. Confirming that the TPS output is near-complete by a full-blown manual search of the entire Q1 database ($\sim$156,000 objects), we ended up with a list of 2680 new EB candidates. This list included transits and eclipses indiscriminately since there is no ready way to distinguish the two. We thus cross-correlated the list of candidate EBs with the list of Kepler exoplanet candidates and removed the overlapping targets. We further conducted an all-hands search for other potential exoplanet candidates, removed them from the list and passed them to the Science Office for further vetting and ground-based follow-up. Inspection of pre-release Q2 and Q3 pipeline products (TPS and Data Validation) contained longer baseline TCEs that yielded an additional $\sim$300 EB candidates.

Other than TPS, we used another automated approach to identify plausible EBs, which entails morphological classification of each light curve's power spectrum. When the power spectrum contains features above a designated power threshold, the spacing of the features was found to be a useful diagnostic for identifying and classifying eclipsing systems. Detached systems have a strong peak near twice the true orbital frequency, followed by a number (several tens in some cases) of higher frequency harmonics at integer or half-integer ratios of the primary frequency. Contact systems, on the other hand, typically have only a few equally spaced strong harmonics. Many non-EB variables could be identified by the lack of equally spaced harmonics in the power spectrum. This method works well for algorithmically identifying EBs with strong features but is less successful as the eclipses become shallower.  Lowering the threshold resulted in larger numbers of non-EB systems being incorrectly selected. Since most candidates were detected by other methods, using this approach was used primarily as validation.

All detected candidates are merged into a master database. In total, 3403 candidate EBs were detected in the \Kepler time-series.

\subsection{Ephemerides}

After initial detection and elimination of duplicates, the ephemerides for all EB candidates have been determined and light curves phased. We limited ourselves to the Q0/Q1 data and we used several approaches in parallel to achieve this goal.

We devised a manual computer tool \emph{ephem} written in python that computes 
periodograms using 3 methods: Lomb-Scargle \citep[LS;][]{lomb1976,scargle1982}, 
Analysis of Variance \citep[AoV;][]{aov1989}, and Box-fitting Least Squares \citep[BLS;][]{kovacs2002}, as implemented in the \verb|vartools| package \citep{hartman2008}. Of the three, BLS performed best, although half-periods were very common. \emph{Ephem} features two panels, one for the periodogram and the other for a phased light curve. Dragging the mouse across the periodogram modifies the period in real time and allows for quick and accurate tuning. Dragging the mouse across the phase plot modifies BJD${}_0$; this way we could set the deeper eclipse to coincide with phase 0.0. 

Fluctuations in the light curves are not always due to the binary nature of the stars (e.g.~star spots, pulsations) and the amplitudes of this ``noise'' can be a sizeable fraction of the eclipse depth. Eclipses can be relatively sharp features in the light curve compared to these effects, and despite it being obvious to the eye, the eclipse signals are in some sense sparse. Thus temporal domain methods (e.g. phase folding) and frequency domain methods (e.g. power spectra) can have difficulty picking out the binary star signal from the other signals in the time series. To circumvent this limitation, and to validate estimates of the period and epoch, we developed a simple tool \emph{sahara} to visually inspect each light curve and select by eye the primary and secondary eclipses via mouse clicks. (For precision, the secondary eclipse is chosen many cycles away.) The code then phase folds the data on the trial period and epoch, and computes a revised period using the minimum string-length method \citep{dworetsky1983}. In most cases the string-length method provides a more accurate period, but for a sizable fraction of cases where spots dominate the power, the by-eye method is clearly superior. A three-panel figure showing the light curve with eclipses marked, the phase-folded light curve, and a close-up of primary eclipse is generated and visually inspected to verify quality of the fiducial epoch (BJD${}_0$) and the period, and if necessary, the process is repeated.

All ephemerides have been manually vetted, but for certain systems it was impossible to uniquely determine the periods, i.e.~for shallow systems with equal depth eclipses (vs.~a single eclipse at half period), or long period systems that feature a single eclipse in the Q0/Q1 data. The accuracy of the determined ephemerides depends on the period; for short-period systems a typical accuracy is $\sim 10^{-5}$, while the periods of tens of days are notably less accurate. We note that both the \emph{ephem} and \emph{sahara} tools are released as open source to any interested parties from \verb|http://phoebe.fiz.uni-lj.si|.

\subsection{Culling}

Phasing of light curves allowed us to manually inspect every target and eliminate false positives. About 27\% of the sample turned out to be non EBs, with offending light curves corresponding mostly to RR Lyr-, $\gamma$-Dor-, $\delta$-Sct and RS CVn-type stars. These have been culled from the sample.

A significant fraction (7\%) of culled elements was due to the artifacts of the eclipsing binary with $P=1.68991$-d that was used as a guide star. As soon as its EB nature was identified, the star was no longer used as a guide star, but systematics in Q0 and Q1 raw data are still present.

The size of the \Kepler pixels projected onto the sky (4 arcsec) coupled with the high star density near the galactic plane lead to a non-negligible likelihood of associating an EB event with the wrong star.  This occurs when the flux of a nearby star (e.g. small angular separation on the sky) impinges on the photometric aperture of the target star in question.  Any variability of the nearby star -- an eclipse signature, for example -- can contaminate the target star light curve, the degree of which depends on the amplitude of variability, the relative magnitude of the stars in question, and their separation on the sky.  This scenario is a common source of false-positives for exoplanet transit detection as described in \citet{batalha2010b} and applies to EB detection as well, though less frequently.  

\begin{figure}
\begin{center}
\includegraphics[width=0.75\textwidth]{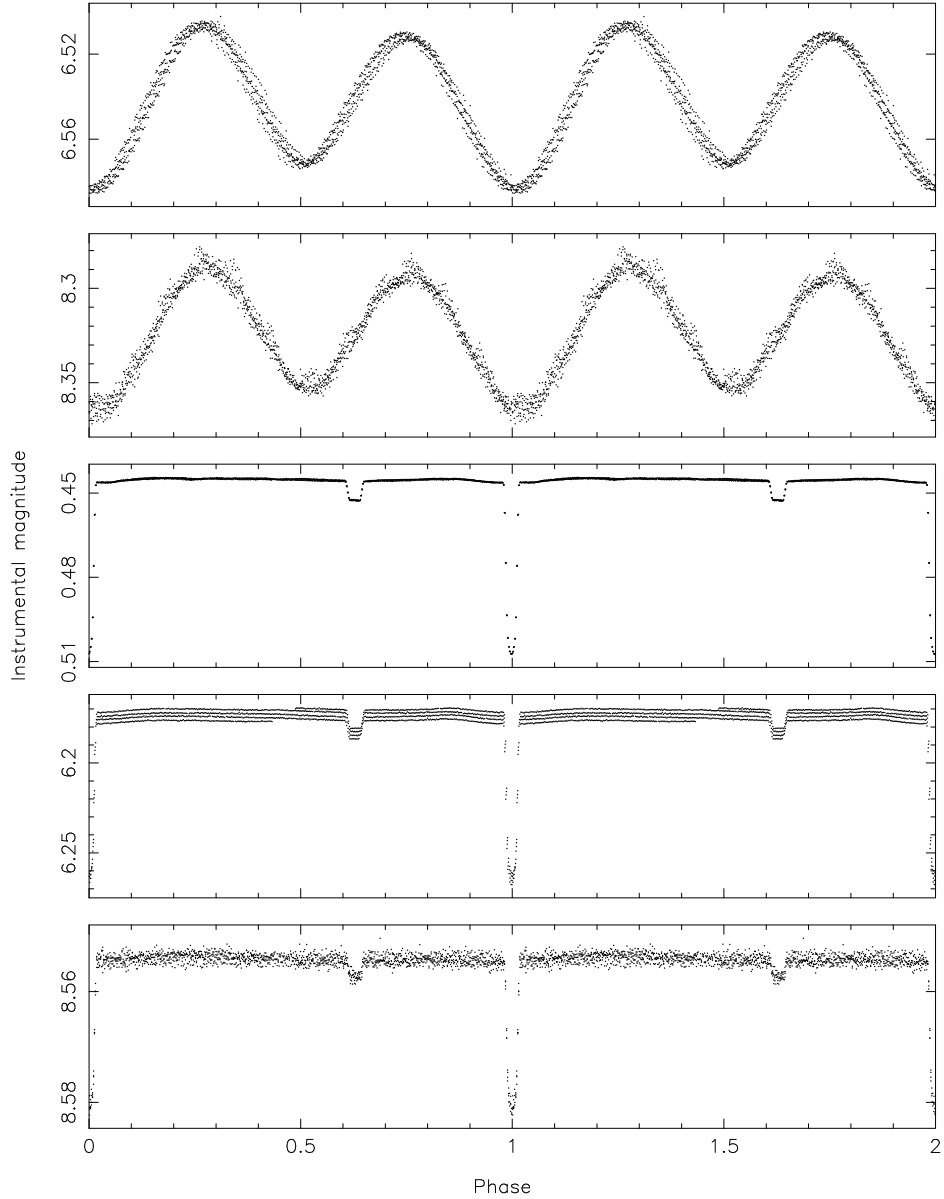} \\
\caption{Examples of spurious detections of contaminating EBs. The $y$-axis is an arbitrary instrumental magnitude scale given by $y=-2.5\log (\mathrm{counts})+26.0$.  The top two stars and their ephemerides are KID 7546791 ($T_0=54964.6060$, $P=0.24231$) and KID 7546789 ($T_0=54964.6023$, $P=0.24233$), respectively.  KID 7546791 is most likely the true binary since it is about 1.75 mag brighter than KID 7546789. The bottom three stars are KID 9851142  ($T_0=54968.8631$, $P=8.48517$), KID 9851126 ($T_0=54968.8573$, $P=8.48516$), and KID 9851123 ($T_0=54968.8558$, $P=8.48027$), respectively. The likely true binary in this case is KID 9851142. There were 61 occurences in the sample that were detected and removed.}
\label{fig_contamination}
\end{center}
\end{figure}

Since \Kepler sends down pixels associated with a small percentage of stars in the field of view, it is generally the rule that the nearby star is not an observed target.  There are exceptions to this, however. Sixty-one spurious EB identifications were removed from the catalog after searching for entries with orbital periods differing by less than 0.1 days and coordinates differing by less than 0.03 degrees.  In each case, the EB identification was assigned to the star whose light curve demonstrated the deepest eclipse event as would be
expected when the photometric aperture is centered on the EB. Fig.~\ref{fig_contamination} shows two examples of contaminated stars. KID 7546791 and KID 7546789 are most likely the same star. Likewise, KID 9851142, KID 9851126, and KID 9851123 are likely to be cross-contaminated.

Tracking the photocenter of the aperture given {\em Kepler}'s very high SNR and stable pointing is an effective means of verifying the source of the EB signal. A flux change in any object whose point spread function falls within the photometric aperture will shift the photocenter of the light distribution.  The Data Validation pipeline examines the photocenter time series of each planet candidate and constructs a motion detection statistic.  Eclipsing binary light curves, however, are not subjected to this analysis, and the pixel-level data are not yet available at MAST.

To compensate for this deficiency, we provide a measure of the flux contamination associated with the photometric aperture of each star.  Column 9 of Table~\ref{table_example} gives the fraction of the total flux in the optimal aperture due to all sources other than the target star itself.  The optimal aperture is defined as the set of pixels that optimizes the total SNR of the flux time-series.  It is dependent on the local Pixel Response Function (PRF), measured on-orbit during the commissioning period \citep{bryson2010}, as well as the distribution of stellar flux on the sky near the target.  The latter relies on information from the {\em KIC}.  An optimal aperture and contamination metric is computed for every star brighter than $Kp=18$ in the {\em KIC}. The EB interpretation should be taken with extreme caution for stars that have a high fraction of flux contamination. Stars with shallow eclipse events should also be regarded with caution even if the flux contamination is modest.

Once culling was completed, the sample was reduced to 1879 EB stars. Note that culling was done somewhat conservatively: if there was an indication that the object might not be a bona fide binary, we tagged it as uncertain. Despite our best efforts, some offenders are bound to still contaminate the sample. As future data become available, the detection, phasing and culling process will be revisited and the longer baseline of observations will provide the critical handle to adequately revise the EB catalog. In addition to the 1879 stars published here, another 21 EBs have been detected but are being held back because their data are still proprietary. These will be added to the catalog in the first revision.

To estimate the completeness of the sample, we conducted a second pass search on a subsample of KID numbers ranging from 9,000,000 to 10,000,000, where all targets have been scrutinously analyzed. There are 19,259 stars in that KID range, out of which 254 are confirmed EBs -- constituting 1.32\%. If we crudely extrapolate this number over the whole \Kepler field to obtain an expected order of magnitude, we get 2058 EBs. This number will vary with EB distribution as function of galactic latitude, but it provides us with the rough quantitative estimate. Compared to the actually detected number of 1879 EBs, this implies that the catalog should be $\geq$91\% complete. As yet another test, we compared the detected sample with the output from the automated classifier by \citet{blomme2010} and found no additional candidates.

\subsection{Classification}

\emph{Detached} binaries are systems in which the separation of both components is large compared to their radii. The stars interact gravitationally, but the distortion of their surfaces due to tidal deformation and rotation is limited. The light curves feature sharp eclipses, and (in the absence of intrinsic stellar oscillations) flat out-of-eclipse regions. Traditionally these are referred to as Algol binaries or EA-type stars. In \emph{semi-detached} binaries one of the stars fills the critical Roche lobe and often features mass transfer. The other component is in a detached state. The eclipses are wide yet still pronounced, out-of-eclipse regions are rounded. Typical representatives are $\beta$ Lyr-type stars. The components in \emph{overcontact} binaries are so close that they share a common envelope. Most commonly known representatives are stars of the W UMa type. The light curves are changing continuously and the ingress and egress points of the eclipses are no longer pronounced. Whenever stellar surfaces are significantly distorted, the visible cross-section varies with phase, causing \emph{ellipsoidal variations}. In the absence of eclipses (low inclination systems) this effect is revealed by the near-sinusoidal light curves.

\begin{figure}
\begin{center}
\includegraphics[width=0.9\textwidth]{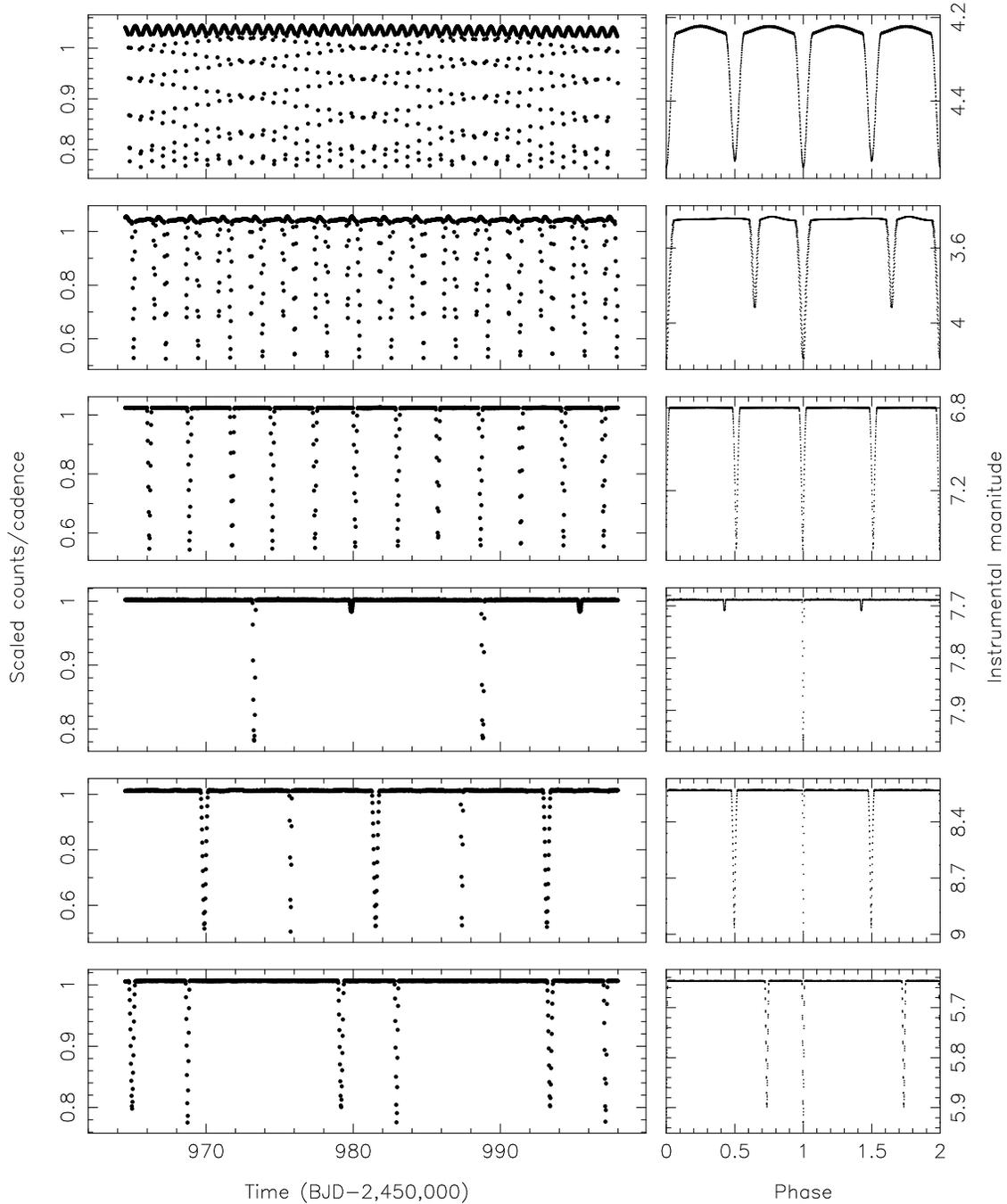} \\
\caption{Example light curves showing EBs classified as detached (D).  The left hand panels show the data in normalized flux units plotted as a function of time.  The right hand panels show the phased light curves in arbitrary magnitude units as in Fig.~\ref{fig_contamination}.  Two binary cycles are shown for clarity.  The stars are, from top to bottom:  KID 5513861 ($P=1.51012$ d), KID 4544587 ($P=2.19074$ d), KID 4445630 ($P=5.62746$ d), KID 6841577 ($P=15.5376$ d), KID 5955321 ($P=11.6347$ d), and KID 9509207 ($P=14.1991$ d).}
\label{plot_det1}
\end{center}
\end{figure}

\begin{figure}
\begin{center}
\includegraphics[width=\textwidth]{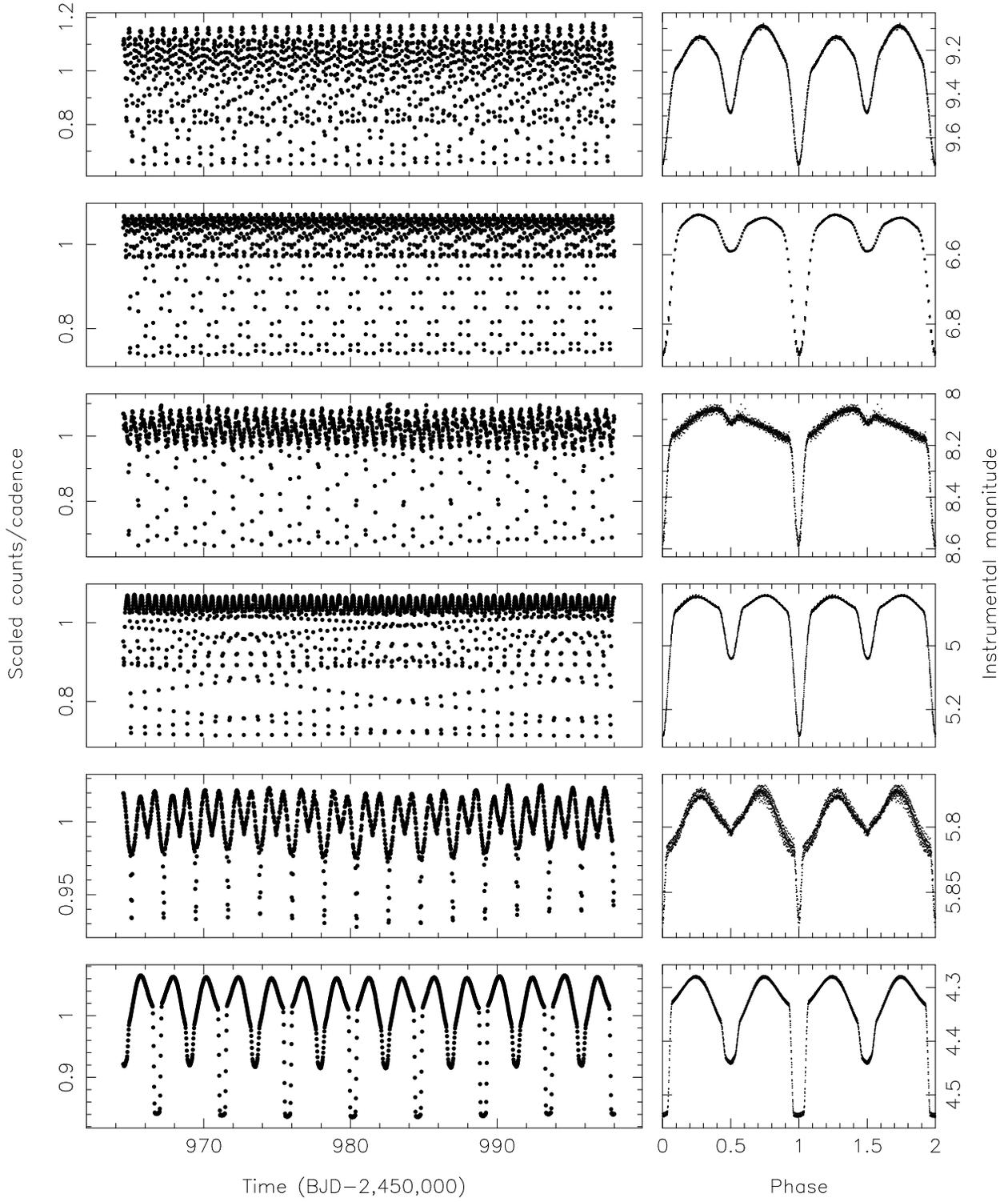} \\
\caption{Similar to Figure \protect{\ref{plot_det1}} for EBs classified as semi-detached (SD).
The stars are, from top to bottom: KID 3218683, ($P=0.77167$ d), KID 8074045, ($P=0.53638$ d), KID 9328852, ($P=0.64581$ d), KID 4729553, ($P=0.96131$ d), KID 11175495, ($P=2.19126$ d), and
KID 8868650, ($P=4.44741$ d).}
\label{plot_sd1}
\end{center}
\end{figure}

\begin{figure}
\begin{center}
\includegraphics[width=\textwidth]{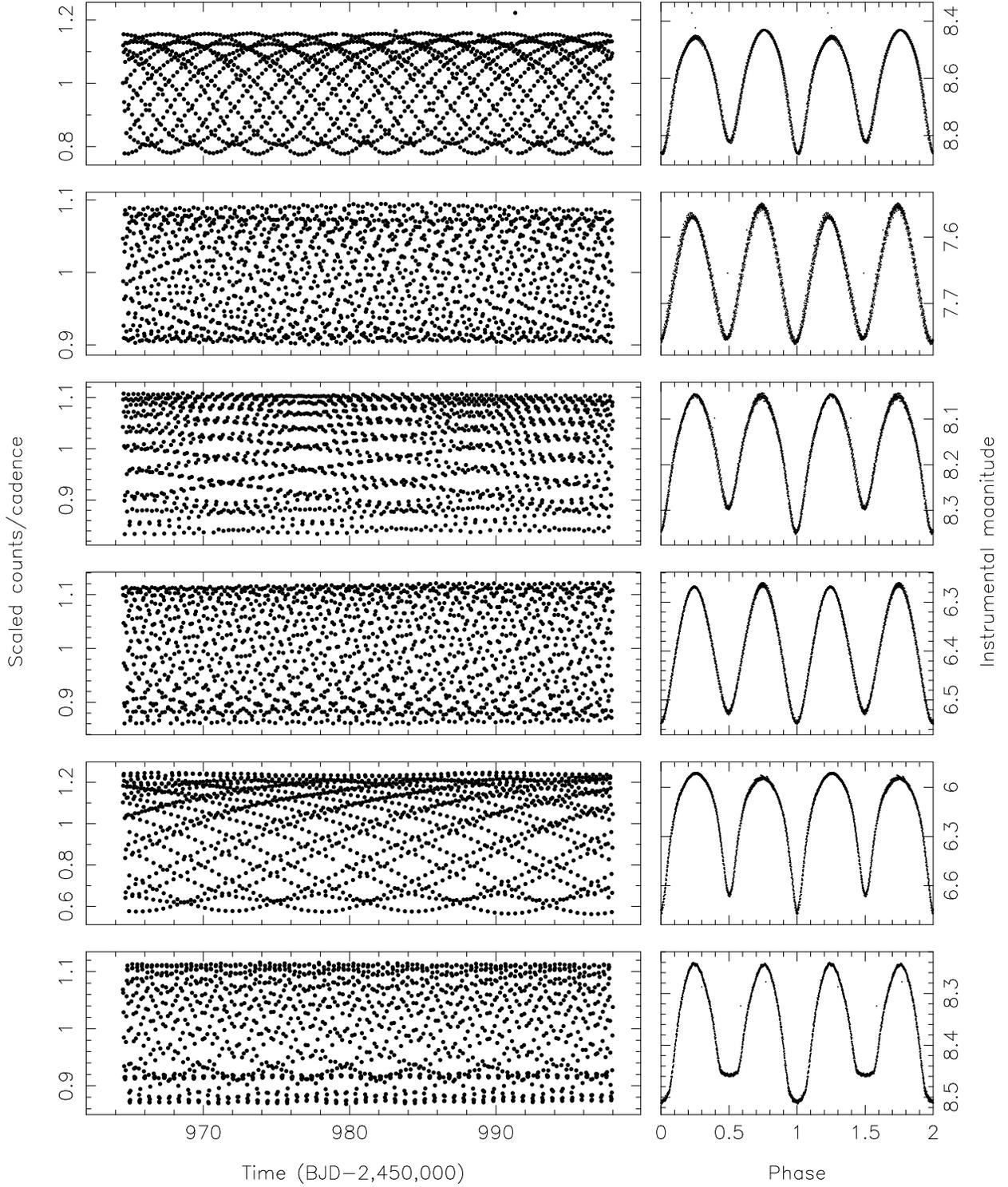} \\
\caption{Similar to Figure \protect{\ref{plot_det1}} for EBs classified as over-contact (OC).
The stars are, from top to bottom:
KID 6106771,  ($P=0.26384$ d),
KID 6265720,  ($P=0.31237$ d),
KID 10447902, ($P=0.33745$ d),
KID 3732732,  ($P=0.39617$ d),
KID 6671698,  ($P=0.47153$ d), and
KID 3127873,  ($P=0.67146$ d).}
\label{plot_oc11}
\end{center}
\end{figure}

\begin{figure}
\begin{center}
\includegraphics[width=\textwidth]{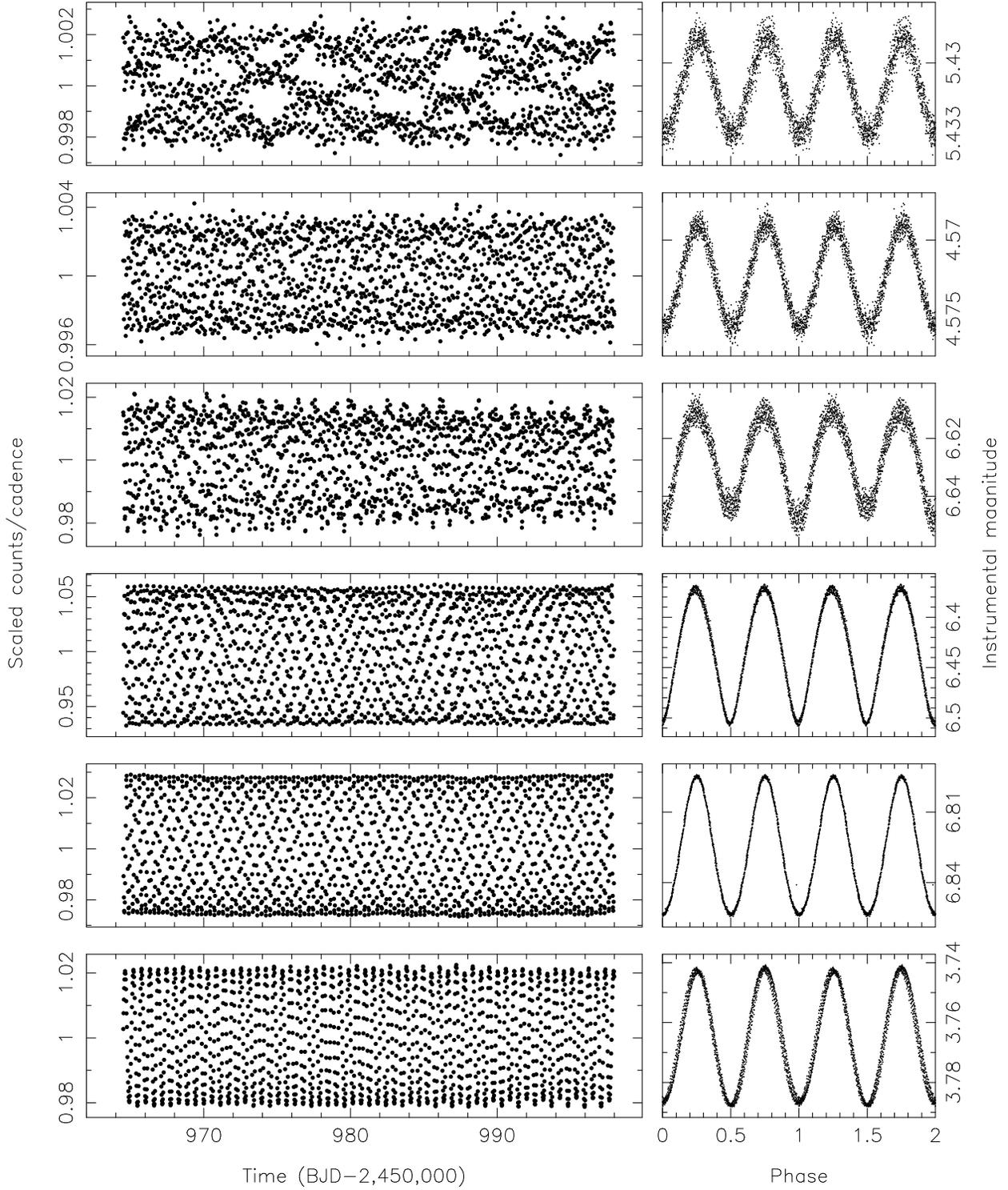} \\
\caption{Similar to Figure \protect{\ref{plot_det1}} for EBs classified as ellipsoidal (ELV).
The stars are, from top to bottom:
KID 9508052,  ($P=0.27999$ d),
KID 4661397,  ($P=0.292301$ d),
KID 4940217,  ($P=0.37873$ d),
KID 9071104,  ($P=0.385229$ d), 
KID 10389982, ($P=0.443392$ d), and
KID 4273411,  ($P=1.21973$ d).
}
\label{plot_ell1}
\end{center}
\end{figure}

\begin{figure}
\begin{center}
\includegraphics[width=\textwidth]{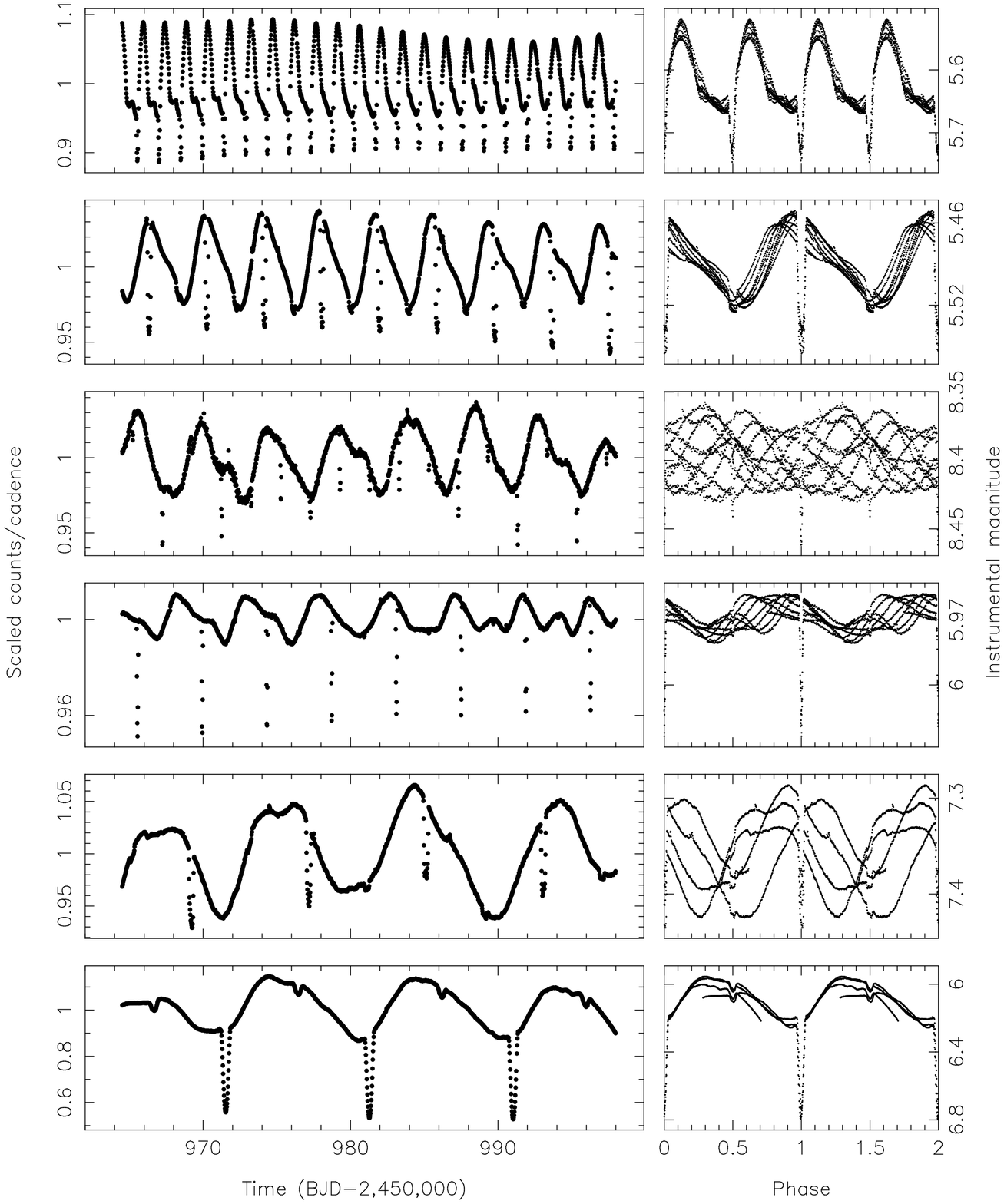} \\
\caption{Similar to Figure \protect{\ref{plot_det1}} for intrinsically variable EBs.
The stars are, from top to bottom:
KID 9137992,  ($P=2.94213$ d),
KID 7940533,  ($P=3.90567$ d),
KID 3427776, ($P=4.01626$ d),
KID 5768927,  ($P=4.39063$ d),
KID 9576197,  ($P=7.96253$ d), and
KID 6197038,  ($P=9.77901$ d).
}
\label{plot_irr1}
\end{center}
\end{figure}

Preliminary classification was done manually by visual target inspection. We classified binaries into 5 groups: detached (D), semi-detached (SD), overcontact (OC), ellipsoidal (ELV), and uncertain (?). The ellipsoidal variable class (ELV) is limited to those systems that exhibit only sinusoidal variations and is not used to indicate a detached system that shows a distinct ellipsoidal effect between eclipses (although there are many that do). Figs.~\ref{plot_det1}-\ref{plot_irr1} depict \Kepler light curves of different morphology classes. These figures showcase the exceptional quality of the \Kepler data.

There are inherent difficulties with this classification, most notably to distinguish over-contact stars and ellipsoidal variables, both of which feature sinusoidal light curves and shallow minima. Furthermore, semi-detached binaries are impossible to classify to high fidelity without accurate modeling, so ``SD'' was used to tag all targets that exhibit large ellipsoidal variations and still feature wide distinct minima. To aid with the first degeneracy, some feedback comes from subsequent modeling (cf.~\S\ref{principal}) through the inclination--fillout factor cross-section for systems that are not severely contaminated with third light. Other attempts have been made, such as the automated classification used for the ASAS project \citep{pojmanski2002}, but the class distinction was not conclusive and manual inspection proved more reliable. The reader should be very cautious not to rely heavily on classification results, especially for short-period objects, as these can potentially change in the future with more data becoming available.

This preliminary manual classification yielded 52.3\% detached EBs, 7.5\% semi-detached EBs, 25.4\% overcontact EBs, 7.5\% ellipsoidal variables, and 7.3\% uncertain types. It is instructive to compare these distributions with other surveys as it highlights the selection effects of ground based surveys. \citet{paczynski2006}'s ASAS study classified 11099 eclipsing binaries into 3 groups: detached, semi-detached, and overcontact, based on their respective domains in the cross-sections of Fourier coefficients $c_2$ and $c_4$. Their study reported the following fractions: 24.8\% detached, 26.6\% semi-detached and 48.6\% overcontact binaries. \citet{christiansen2008}'s study of 850 variables in the 25 target fields lump detached and semi-detached binaries and report a 37.1\% frequency, while overcontacts constitute 56.9\% of the sample and ellipsoidals constitute 6\% of the sample. It is evident that {\it Kepler} is superiorly sensitive to long(er) period detached binaries, mostly because of the continuous data coverage and unprecedented precision. Another example comes from the OGLE survey that provided a census of 2768 EBs in the Large Magellanic Cloud \citep{wyrzykowski2003}. They used an image recognition neural network to classify EBs into three classes: detached (68.0\%), semi-detached (25.9\%) and overcontact (6.1\%). Here the sample is biased by the luminosity limit towards bright components and thus the deficit in overcontact systems is very pronounced. While Kepler typically does not target hot stars, there are only a few in the field and hence there should be little-to-no luminosity-induced selection effects present in the database.

\subsection{Description of the catalog}

The catalog contains 1879 EBs. For each EB, we provide: its \Kepler ID number (column 1); ephemeris: BJD${}_0$-2400000 (column 2) and period in days (column 3); classified morphology (column 4): "D", "SD", "OC", "ELV" or "?"; \Kepler magnitude\footnote{\Kepler magnitude is computed according to a hierarchical scheme and depends on pre-existing source catalogs: Sloan, 2MASS, Tycho 2 or photographic photometry. Hence the value does not correspond to any given binary phase and provides only a rough idea of the object intensity.} (column 5); input catalog parameters: $T_\mathrm{eff}$ in K (column 6), $\log g$ in cgs units (column 7), $E(B-V)$ (column 8); crowding (column 9); principal parameters (viz.~\S\ref{principal}): $T_2/T_1$ (column 10), mass ratio $q$ (column 11), fillout\footnote{$F = (\Omega-\Omega_{\mathrm L2})/(\Omega_\mathrm{L1}-\Omega_\mathrm{L2})$, where $\Omega$ is the gravity potential as defined in \citealt{wilson1979}.} factor $F$ (column 12), the sum of fractional radii $\rho_1+\rho_2 \equiv (R_1+R_2)/a$ (column 13), radial component of eccentricity $e \sin \omega$ (column 14), tangential component of eccentricity $e \cos \omega$ (column 15), and sine of inclination $\sin i$ (column 16). Table \ref{table_example} shows 21 representative entries from the catalog. The table in its entirety is provided as an online supplement to this paper.

Periods cannot be determined reliably for well detached systems that feature a single eclipse in Q0/Q1 data. For those systems we do not provide the minimum period -- these will be updated as more data become available. For those cases the BJD${}_0$ field lists the actual time of minimum.

\begin{table}
\caption{Database of EBs in \Kepler Q0/Q1 data (abridged).} \label{table_example}
\begin{center}
\scriptsize
\begin{tabular}{lrrcccccc}
\hline \hline
KID     &  BJD${}_0$  &  P${}_0$  & Type & K mag & $T_\mathrm{eff}$ & $\log g$ & $E(B-V)$ & crowding \\
        &             & $T_2/T_1$ & $q$  & $F$   & $\rho_1+\rho_2$  & $e \sin \omega$ & $e \cos \omega$ & $\sin i$ \\
\hline
7938468 & 55005.30055 & 7.22693 & D & 13.782 & 5858 & 4.265 & 0.102 & 0.067 \\
&& 0.64775 &&& 0.54478 & 0.01819 & -0.00199 & 1.00182 \\
8345358 & 55003.08302 & 9.38058 & D & 13.27 & 5899 & 4.209 & 0.098 & 0.005 \\
&& 0.86324 &&& 0.10854 & 0.04117 & -0.04851 & 0.99953 \\
8075755 & 54964.75525 & 0.49620 & D & 11.84 & 6153 & 4.258 & 0.067 & 0.017 \\
&& 0.76796 &&& 0.14238 & -0.11244 & -0.05652 & 0.99382 \\
8075618 & 54970.92830 & 17.55980 & D & 12.925 & 6351 & 4.531 & 0.079 & 0.004 \\
&& 0.75247 &&& 0.06932 & -0.30507 & -0.16697 & 1.00518 \\
7938870 & 55002.47550 & 0.58072 & SD & 12.331 & 5941 & 4.444 & 0.062 & 0.006 \\
&& 0.40441 &&& 0.62235 & 0.05635 & -0.05966 & 0.97364 \\
6836140 & 55001.72627 & 0.48772 & SD & 14.212 & 5712 & 4.464 & 0.100 & 0.112 \\
&& 0.86056 &&& 0.63739 & 0.02583 & 0.00346 & 0.95109 \\
8074045 & 55002.51687 & 0.53638 & SD & 15.344 & 5307 & 4.592 & 0.106 & 0.029 \\
&& 0.47882 &&& 0.64713 & -0.01492 & -0.08358 & 0.97806 \\
6669809 & 55005.35890 & 0.73373 & SD & 14.4 & 4144 & 4.541 & 0.038 & 0.012 \\
&& 0.76798 &&& 0.64613 & 0.08553 & 0.00098 & 0.94325 \\
8539720 & 55005.35160 & 0.74448 & OC & 12.295 & 6651 & 4.230 & 0.091 & 0.000 \\
&& 0.89027 & 0.33588 & 0.78990 &&&& 0.84088 \\
8539850 & 54999.41852 & 0.64208 & OC & 15.674 & 5288 & 4.621 & 0.107 & 0.089 \\
&& 1.01396 & 1.33254 & 0.97572 &&&& 0.54026 \\
8143757 & 55001.73880 & 0.35650 & OC & 15.653 & 5317 & 4.533 & 0.117 & 0.000 \\
&& 0.97624 & 1.15554 & 0.17374 &&&& 0.84196 \\
7584739 & 55001.83432 & 0.91155 & OC & 15.235 & 5464 & 4.509 & 0.112 & 0.023 \\
&& 0.85548 & 0.91354 & 0.13734 &&&& 0.82955 \\
7941050 & 55002.62385 & 0.29797 & ELV & 11.541 & 6138 & 4.212 & 0.065 & 0.000 \\
&& 0.93952 & 0.39861 & 1.02008 &&&& 0.77701 \\
7091476 & 55078.77350 & 0.80772 & ELV & 12.226 & 4587 & 2.621 & 0.135 & 0.004 \\
&& 0.99660 & 1.20847 & 0.99589 &&&& 0.57935 \\
7936219 & 55002.58874 & 0.38560 & ELV & 12.373 & 8848 & 3.843 & 0.148 & 0.009 \\
&& 1.03169 & 1.74968 & 0.85135 &&&& 0.44968 \\
6844489 & 54965.37350 & 1.07980 & ELV & 11.9 & 6150 & 4.231 & 0.071 & 0.004 \\
&& 0.93275 & 0.71850 & 0.26758 &&&& 0.83978 \\
8280135 & 54965.65265 & 0.28689 & ? & 14.596 & 3648 & 4.624 & 0.022 & 0.779 \\
&& &&& & & & \\
7176440 & 55002.54692 & 0.35840 & ? & 15.821 & 5247 & 4.715 & 0.108 & 0.089 \\
&& &&& & & & \\
5517211 & 55018.78564 & 0.45868 & ? & 14.549 & 5366 & 4.933 & 0.066 & 0.024 \\
&& &&& & & & \\
6026204 & 55001.86208 & 2.28209 & ? & 11.496 & 3185 & 0.046 & 0.147 & 0.028 \\
&& &&& & & & \\
\hline
\end{tabular}
\normalsize
\end{center}
\end{table}

\section{Principal parameters of the sample} \label{principal}

To determine statistical properties of the sample, we employed the EBAI method (Eclipsing Binaries via Artificial Intelligence; \citealt{prsa2008}). The method relies on trained neural networks to yield principal parameters for every binary in the sample.

\subsection{The training set}

The morphology of EBs determines the set of principal parameters that can be extracted from a single light curve. There are useful symmetries that we can employ for overcontact binaries, most notably the same equipotential $\Omega$ for the common envelope. Ellipsoidal variations in a light curve hint on the mass ratio; for total eclipses, the mass ratio can be determined reliably. Eclipse shape determines the degree of contact. Because of proximity, these stars have long circularized. Yet geometric contact does not necessarily imply thermal contact -- in fact, more than 40\% of all overcontact binaries have temperature ratios $T_2/T_1 \leq 0.95$ \citep{pilecki2009}.

In semi-detached binaries, one star fills the Roche lobe exactly and is thus fairly well constrained, but the other star is geometrically detached and can have any radius and effective temperature. Except for the total eclipsers, the photometric mass ratio for these systems is only marginally constrained. Semi-detached systems typically have circularized orbits.

Detached EB light curves are the most complex. Except for the tightest of systems, they contain no information on the mass ratio or on any absolute scale. The eccentricity is often non-zero, and stellar evolution of both components is only loosely coupled.

To account for these fundamental differences, we created two distinct neural networks, one for detached and semi-detached EBs and the other for overcontacts, with principal parameters summarized in Table \ref{table_nn_pars}. For overcontact binaries, we selected four principal parameters: temperature ratio $T_2/T_1$, photometric mass ratio $q_\mathrm{ph}$, the fillout factor $F$, and the sine of inclination. For semi-detached and detached binaries the role of $F$ is superceded by the sum of fractional radii $\rho_1+\rho_2$. For detached binaries there is no handle on $q_\mathrm{ph}$, and we account for eccentricity $e$ and argument of periastron $\omega$ in their orthogonalized forms $e \sin \omega$ and $e \cos \omega$.

\begin{table}
\caption{Principal parameters that correspond to distinct morphology types: overcontact (OC); detached and semi-detached (D/SD).} \label{table_nn_pars}
\begin{center}
\begin{tabular}{cl}
\hline \hline
Morphology: & Parameters: \\
\hline
OC          & $T_2/T_1$, $q_\mathrm{ph}$, $F$, $\sin i$ \\
D/SD        & $T_2/T_1$, $\rho_1+\rho_2$, $e \sin \omega$, $e \cos \omega$, $\sin i$ \\
\hline
\end{tabular}
\end{center}
\end{table}

\begin{figure}
\begin{center}
\includegraphics[height=\textwidth,angle=-90]{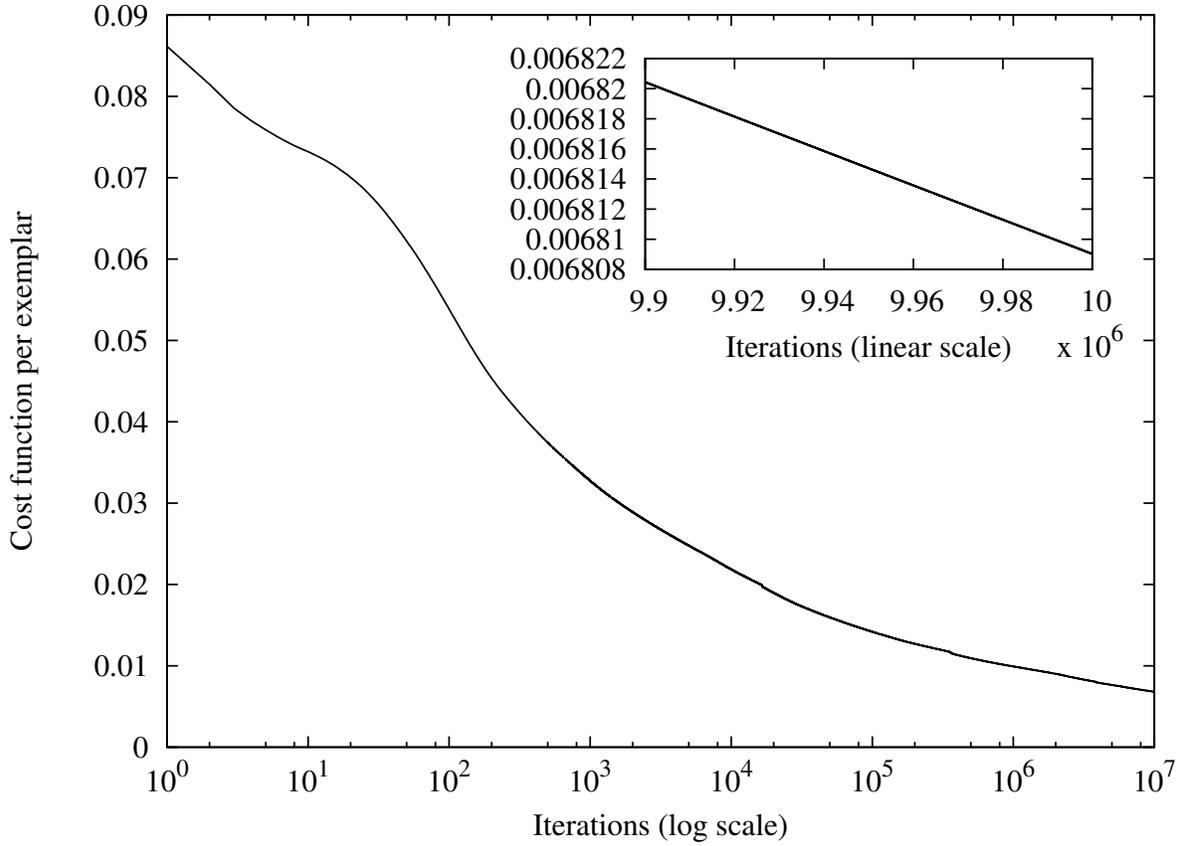} \\
\caption{The neural network learning curve for overcontact binaries. The cost function is defined as the average deviation per parameter per exemplar. The data on the main plot are depicted in log scale, while the data in the inset are in linear scale. The network was trained on 10,000 exemplars for 10,000,000 iterations.} \label{fig_learning}
\end{center}
\end{figure}

The neural network for overcontact EBs is trained on a synthetic sample of 10,000 
stars (exemplars) generated by the light curve synthesis code PHOEBE \citep{prsa2005}. The principal parameters of light curves in the sample are drawn randomly from the probability distribution functions (PDFs) that describe physically plausible systems. These PDFs are carefully selected to optimize the performance of the neural network. Each light curve is synthesized using the \Kepler passband transmission function. Based on $T_\mathrm{eff}$ and $\log g$, the limb darkening coefficients are obtained by interpolation from the updated lookup tables; canonical values for gravity brightening coefficients (1.0 and 0.32) and albedoes (1.0 and 0.6) for radiative and convective envelopes were used, respectively. To each light curve we added variable jitter between 0.05\% and 5\% to aid the networks' recognition capabilities (cf.~\citealt{prsa2008} for a detailed discussion on the training strategy). The training was done in 10,000,000 iterations with a parallelized version of the back-propagation code on a 24-node Beowulf cluster. Fig.~\ref{fig_learning} depicts the learning curve (average deviation per parameter per exemplar as function of iteration). Training took $\sim$4 days to complete.

The detached EB network is trained on a synthetic sample of 35,000 exemplars. The increased number serves to adequately cover the much more complex parameter space of detached EBs. The procedure for creating the training sample is identical in all other aspects to that for overcontact EBs.

As input, neural networks take a set of equidistant phase points. To achieve this, observations are pre-processed with \emph{polyfit}, a polynomial chain fitter \citep{prsa2008}, to obtain a suitable analytical representation that can then be evaluated in an equidistant set of phases. Although synthetic light curves can readily be computed equidistantly (and they usually are), they are also pre-processed with \emph{polyfit} to improve network recognition and to mitigate systematics that stem from any polyfit artifacts. This way the network is matching the same type of data. Although substantial care was taken to validate the ephemerides, \emph{polyfit} automatically determines the primary eclipse minimum by least squares fitting and shifts the phased light curve so that the minimum appears exactly at phase 0.

\subsection{EBAI validation}

The performance of trained neural networks was validated on two distinct sets of 10,000 synthetic light curves generated with PHOEBE, according to the same PDFs as the training sample. These sets were generated only to test the network's ability to yield reliable parameters, they were not used to train the network. The results of validation are presented in Figs.~\ref{fig_test_set_results} and \ref{fig_test_set_diffs} (overcontacts) and Fig.~\ref{fig_d_test_set_results} (detached EBs).

\subsubsection{EBAI performance on overcontact EBs.}

Principal parameters of overcontact EBs have been determined to better than 5\% in the following fractions: $T_2/T_1$ for 94\%, $q$ for 80\%, $F$ for 51\% and $\sin i$ for 97\% of the sample. We note the most significant recognition features below.

\begin{figure}
\begin{center}
\includegraphics[height=\textwidth,angle=-90]{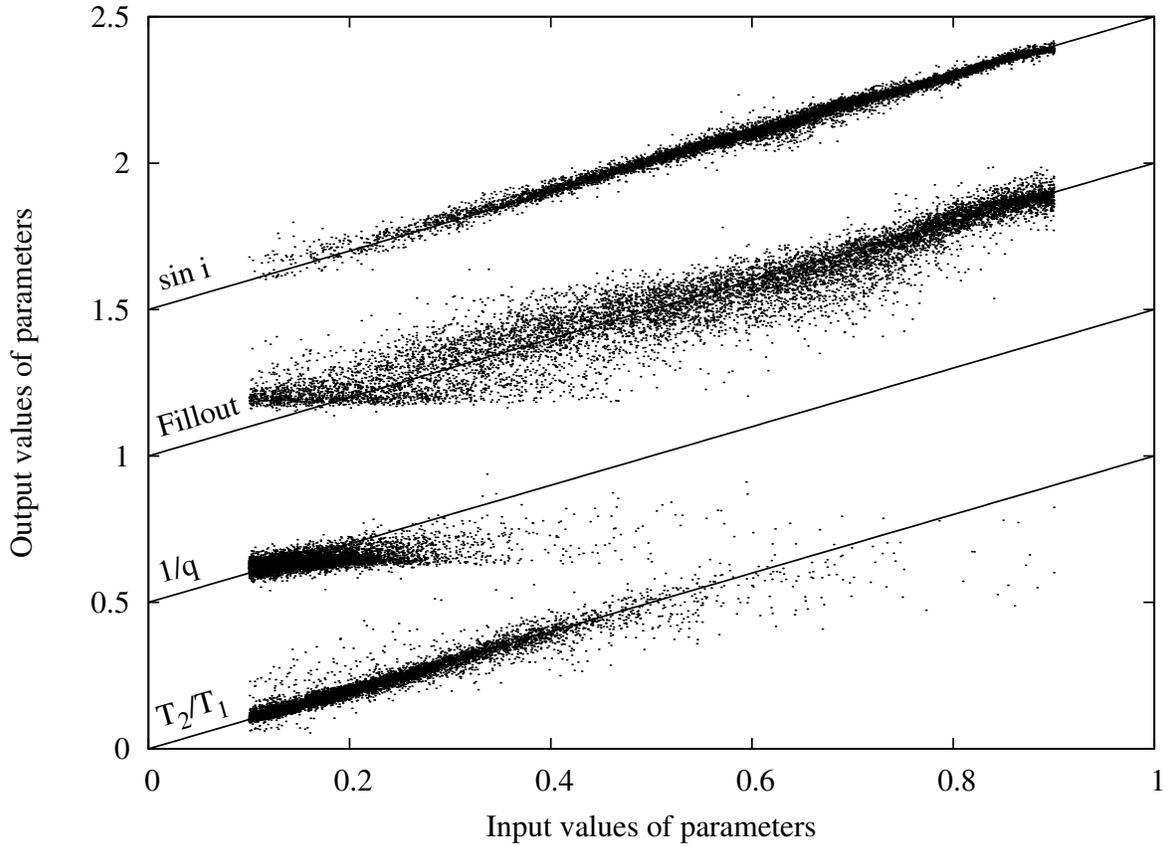} \\
\caption{EBAI performance on a test set of 10,000 overcontact EBs distinct from the training sample. The correlation between input values and recognized output values is shown for each parameter and offset by 0.5 for clarity. Labeled guidelines representing ideal performance are provided for easier comparison. The observed trends and areas of increased scatter are well understood and discussed in the text.} \label{fig_test_set_results}
\end{center}
\end{figure}

\begin{figure}
\begin{center}
\includegraphics[height=\textwidth,angle=-90]{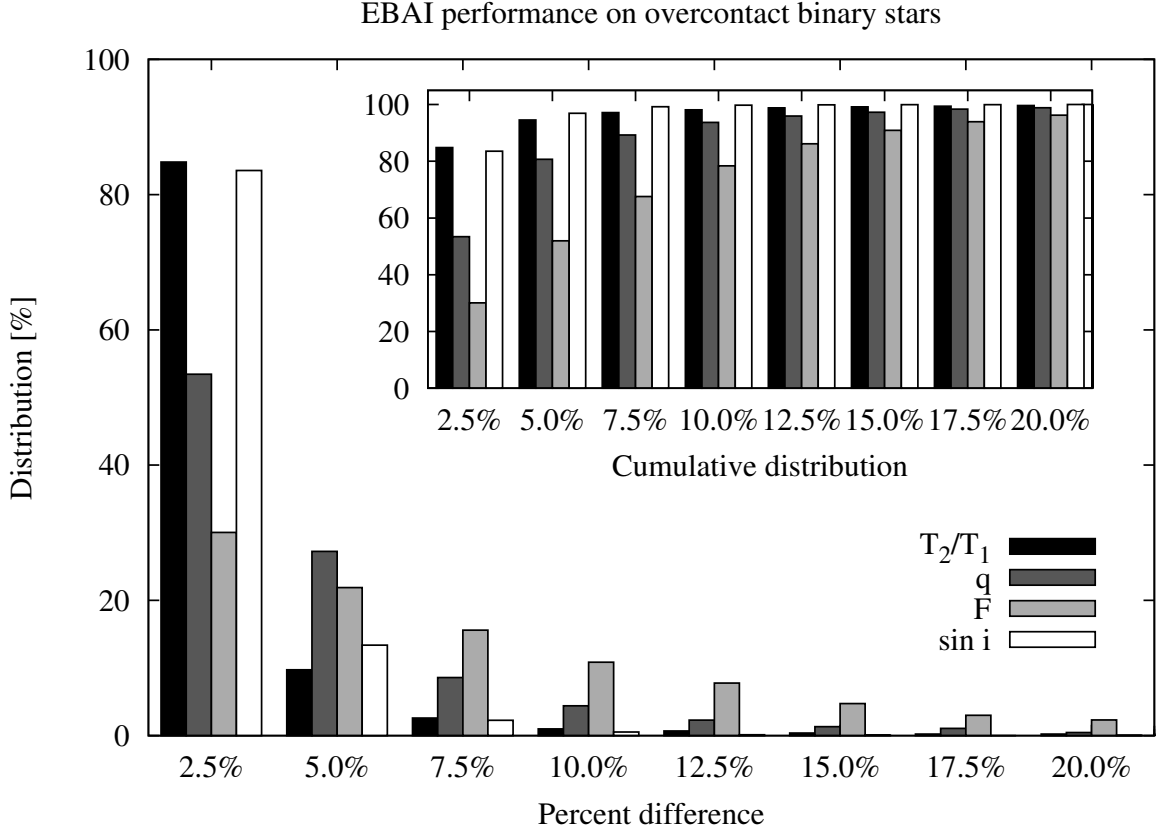} \\
\caption{The distributions of differences between the test set and the network output for overcontact binaries. The inset depicts the cumulative distributions. While the inclination and the temperature ratio are remarkably well determined, the mass ratio is determined reliably only for high inclination systems, and fillout factor is quite dispersed. Nevertheless, 90\% of all systems have errors in all 4 parameters smaller than 15\%.} \label{fig_test_set_diffs}
\end{center}
\end{figure}

\begin{figure}
\begin{center}
\includegraphics[height=\textwidth,angle=-90]{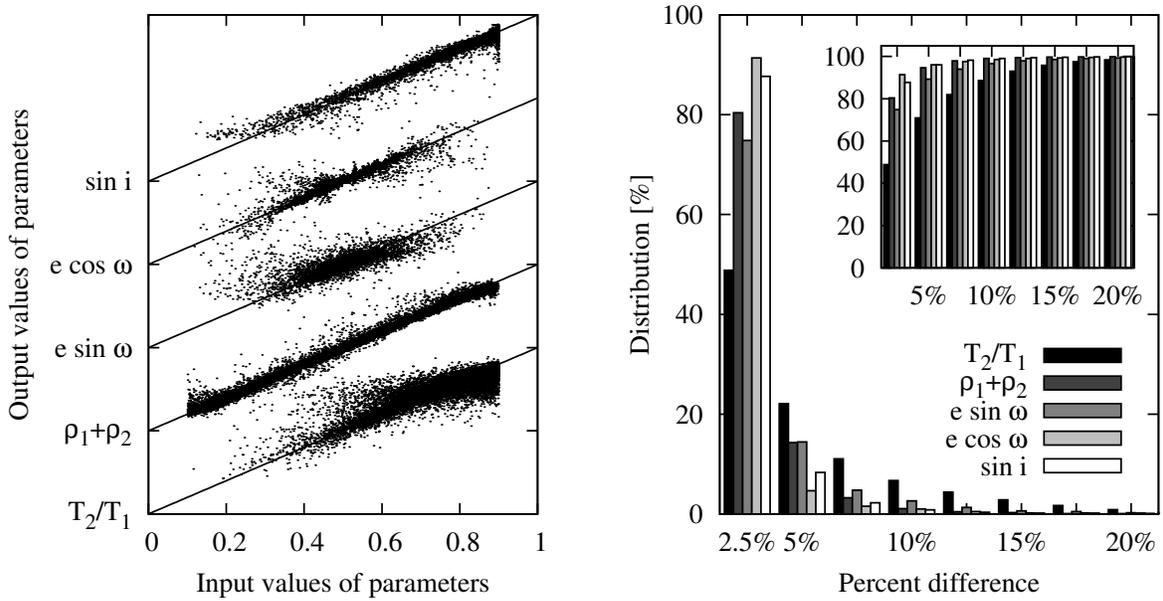} \\
\caption{EBAI performance on a test set of 10,000 detached EBs distinct from the training sample. Despite the increased morphology complexity, the network still yields a statistically significant result: 90\% of all systems have errors in all 5 parameters smaller than 10\%.} \label{fig_d_test_set_results}
\end{center}
\end{figure}

\begin{description}
\item[$T_2/T_1$.] The eclipse depth ratio directly depends on the surface brightness ratio. For stars of similar sizes this is approximated to a high degree by the temperature ratio $T_2/T_1$. It is thus not surprising that this parameter is reproduced very reliably. Scatter increases with decreasing $T_2/T_1$, which is the regime where $T_2/T_1$ becomes a poor proxy for the surface brightness ratio.

\item[$q$.] The mass ratio is traditionally determined from spectroscopic observations. However, this parameter's signature can be detected in photometric light curves as a second order effect: it determines the shape of the common envelope and may be deduced from the ratio of polar radii of both components. For total eclipses, this ratio can readily be obtained and the mass ratio can thus be derived \citep{terrell2005}. For partial eclipses this becomes more difficult, although \Kepler's ultra-high photometric accuracy holds great promise for manual analysis. For ellipsoidal variables this signature is largely lost. The correlation between $q$ and the ratio of equipotential radii is deteriorated with a decreasing value of $q$. The dispersion on Fig.~\ref{fig_test_set_results} attests to these limitations; the network yields reliable mass ratios only for total eclipsers with mass ratios close to 1, and shows strong signs of systematics for other systems. This is not a deficiency of the network as such, but rather the inherent limitation to derive a photometric mass ratio for those systems.

\item[$F$.] The degree of overcontact is used as a proxy to the bounding equipotential $\Omega$. Since $\Omega$ depends explicitly on the mass ratio, and the network performs optimally for orthogonal parameters, we replaced it with $F$ as a principal parameter. $F = 0$ corresponds to stars that are bound by the Roche lobe (critical potential through $L_1$), while $F = 1$ corresponds to stars that are in complete contact, bound by the equipotential through $L_2$. Values smaller than 0 indicate detached systems, while values larger than 1 are unphysical. Similar to the mass ratio, $F$ correlates with the ratio of equipotential sizes, but the correlation does not deteriorate with increasing $F$. That is why the correlation is evident across all values of $F$, although the dispersion is quite large. For systems near coalescence the equipotential size ratio cannot be reliably determined, so the handle on $F$ is lost as well. Moreover, such light curves strongly resemble those of ellipsoidal variables, which is revealed in a correlation between $F$ and $\sin i$.

\item[$\sin i$.] Among the 4 parameters, inclination is determined most reliably. 
This stems from a linear relationship between $\sin i$ and the overall amplitude of the light curve. Accurate reproduction fails only for the lowest inclinations, nearly face-on systems ($i < 10^\circ$). However, the inclination is \emph{critically} correlated with excess light, both foreground and background, and the values will be significantly underestimated with increased crowding. This bears special significance for \Kepler because of the large 4 arcsec pixel size.
\end{description}

\subsubsection{EBAI performance on detached EBs.} \label{d_validation}

Principal parameters of detached EBs have been determined to better than 5\% in the following fractions: $T_2/T_1$ for 71\%, $\rho_1 + \rho_2$ for 95\%, $e \sin \omega$ for 89\%, $e \cos \omega$ for 96\%, and $\sin i$ for 96\% of the sample. As before, we note the most significant recognition features.

\begin{description}
\item[$T_2/T_1$.] While for overcontacts this parameter is well determined, for detached binaries it is not. The scatter is most pronounced near $T_2/T_1 \approx 1$ where equal depth eclipses are expected. However, the eclipse depth ratio is strongly affected by eccentricity and star sizes as well, rendering $T_2/T_1$ a poor proxy to the surface brightness ratio. The scatter for low values of $T_2/T_1$ stems from the same sources as for overcontacts.

\item[$\rho_1+\rho_2$.] The sum of fractional radii is directly related to the width of the eclipse and is thus reproduced very reliably. The fall-off at the lower end is due to the limited phase coverage: sharp eclipses feature only a few phase points and the determination of the eclipse width becomes difficult. The upper end deviation, on the other hand, is due to the degeneracy between the inclination and the radii: a slight increase in the radius can be mimicked by a slight decrease in inclination.

\item[$e \sin \omega$.] The radial component of eccentricity is determined by the eclipse duration ratio. It is thus a second order effect. Whenever this ratio is well determined from photometry, this parameter can be obtained reliably. However, with limited data coverage this is increasingly difficult with the longer time-spans. Additional \Kepler data for later quarters will significantly improve this determination.

\item[$e \cos \omega$.] The tangential component of eccentricity is much better determined because its signature in light curves is directly related to the phase separation of both eclipses. It is thus a first order effect.

\item[$\sin i$.] Similar to overcontacts, the inclination of detached EBs is determined reliably. The mismatch on the grazing end bears little significance because it corresponds to non-eclipsing systems with more-or-less constant light. The deviation on the upper end is due to the already mentioned degeneracy between the inclination and the radii.
\end{description}

\centerline{\S}

This validation demonstrates the capability of the ANN to successfully recognize data it has never encountered before. 90\% of the overcontact sample has errors in all 4 parameters smaller than 15\%; 90\% of the detached sample has errors in all 5 parameters smaller than 10\%. This implies that the network output on unknown data is viable for statistical analysis and as input to sophisticated modeling engines for fine-tuning. However, it should be noted that Kepler data often exhibit features that have not been accounted for in the training samples, most commonly intrinsic variability, chromospheric activity (spots), dynamical perturbations by tertiaries and third light contamination. These effects are bound to introduce additional uncertainty to the determined parameters and we warn against using the solutions indiscriminately. Further refinement will be done using physical models (i.e.~WD or PHOEBE) and published later in this paper series.

\section{Results}

\subsection{Statistical Analysis}

In this section we provide several views into the distribution of eclipsing binaries in the \Kepler FOV.

\begin{figure}
\begin{center}
\includegraphics[height=\textwidth,angle=270]{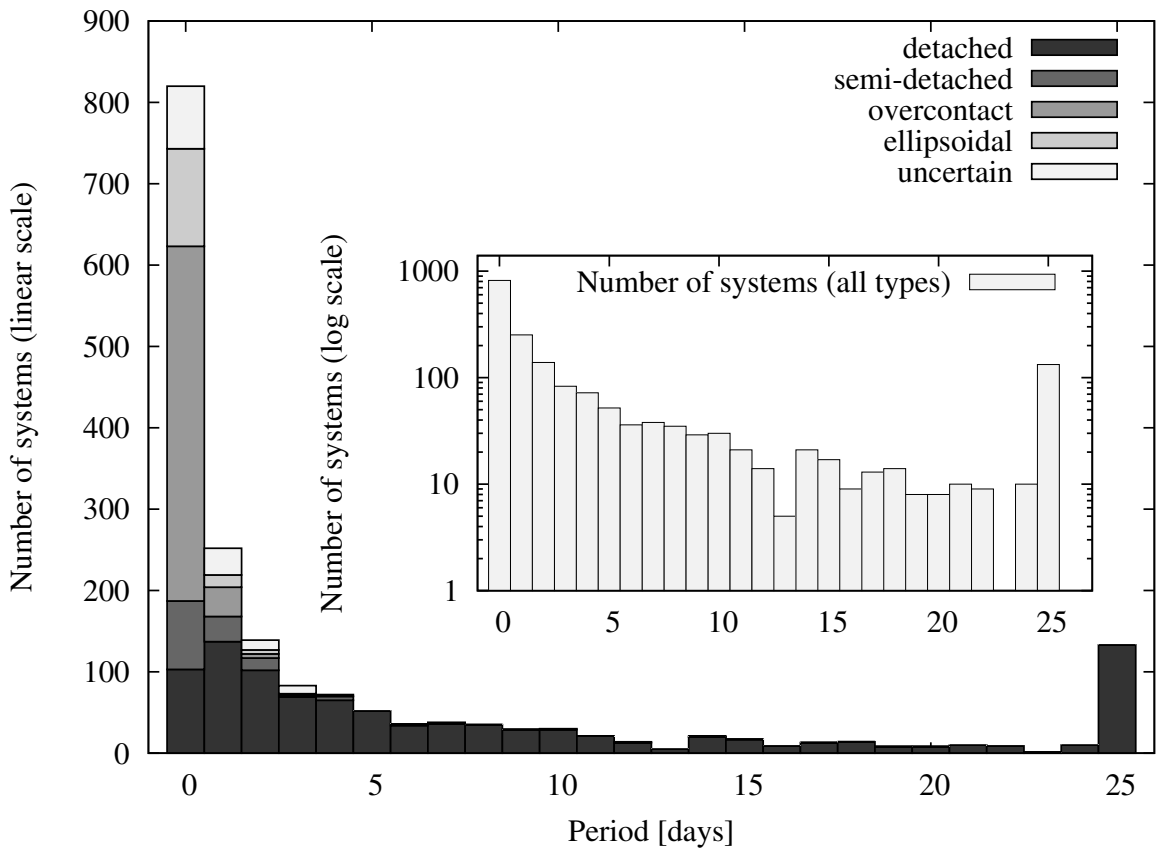} \\
\caption{The distribution of periods for all the stars identified as eclipsing binaries in the \Kepler Q1 data set.  The baseline was 34 days (44 with Q0).  The bars are stacked by morphology types (detached, semi-detached, overcontact, ellipsoidal and uncertain). The inlet depicts the number of systems in log scale.} \label{fig_ebperiods}
\end{center}
\end{figure}

Figure \ref{fig_ebperiods} is a histogram showing the number of EBs as a function of their orbital period. The fraction of detached, semi-detached, overcontact and ellipsoidal systems in each period bin is indicated by the greyscale. The short period excess is attributed to ellipsoidal variables and overcontact binaries. This histogram can be readily compared to the corresponding histogram for planet candidate periods in \citet{borucki2010b}, part 5, Fig.~5.

\begin{figure}
\begin{center}
\includegraphics[height=\textwidth,angle=270]{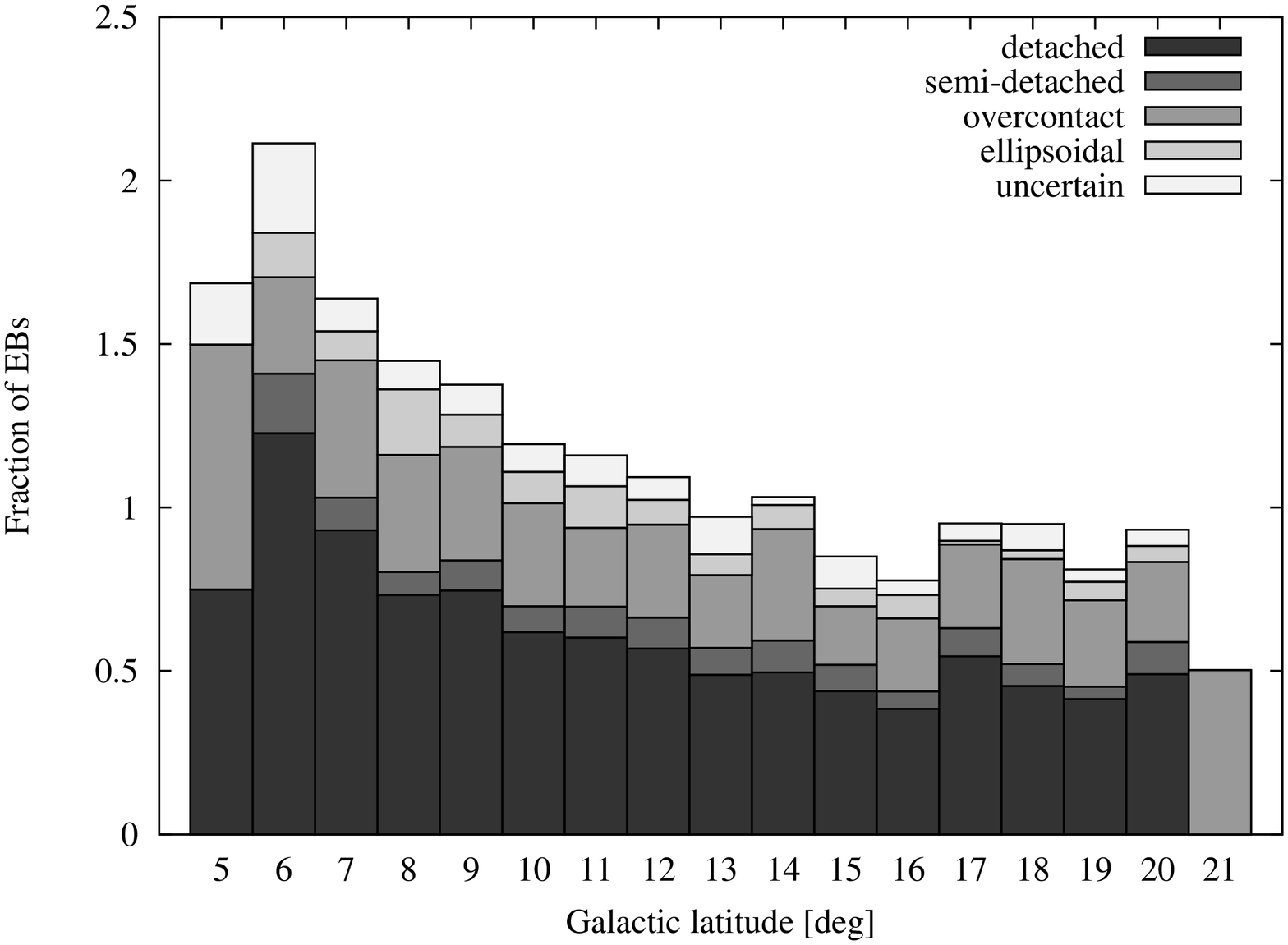} \\
\caption{The distribution of EBs as function of galactic latitude, normalized to the number stars per degree observed by Kepler. Morphology types are denoted with different grayscales. The trend is a consequence of the Kepler target selection.} \label{fig_eblatitude}
\end{center}
\end{figure}

Fig.~\ref{fig_eblatitude} displays the fraction of all stars in the Q1 \Kepler dataset that are included as EBs in this catalogue in strips of 1.0-deg of galactic latitude across the \Kepler FOV. At the higher latitudes the EB fraction is $\sim$1.1--1.2\%, somewhat greater than reported in other surveys. We attribute this to the greater photometric precision of the \Kepler photometer that reveals shallower systems (highly diluted and nearly face-on ellipsoidals) than was previously possible. At the lower galactic latitudes covered by the \Kepler FOV, the evident increase in the fraction of EBs suggests crowding effects are becoming increasingly important. This result bears significance for determining the rate of false positives among planet candidates. Detached binaries with periods of a few days and longer are potential sources of false positives: their light contaminates the observed target by contributing a fraction of the light into the aperture for that target, causing small dips that are often mistaken for planet transit signatures.

\subsection{EBAI results}

The \Kepler sample of EBs was divided into two groups, based on the morphology 
type: 1) overcontact and ellipsoidal, and 2) semi-detached and detached. For the time being, uncertain types were not analyzed. Each group was pre-processed with {\sl polyfit} and submitted to the corresponding neural network. Forward propagation takes less than a second on a single processor.

\begin{figure}
\begin{center}
\includegraphics[height=\textwidth,angle=-90]{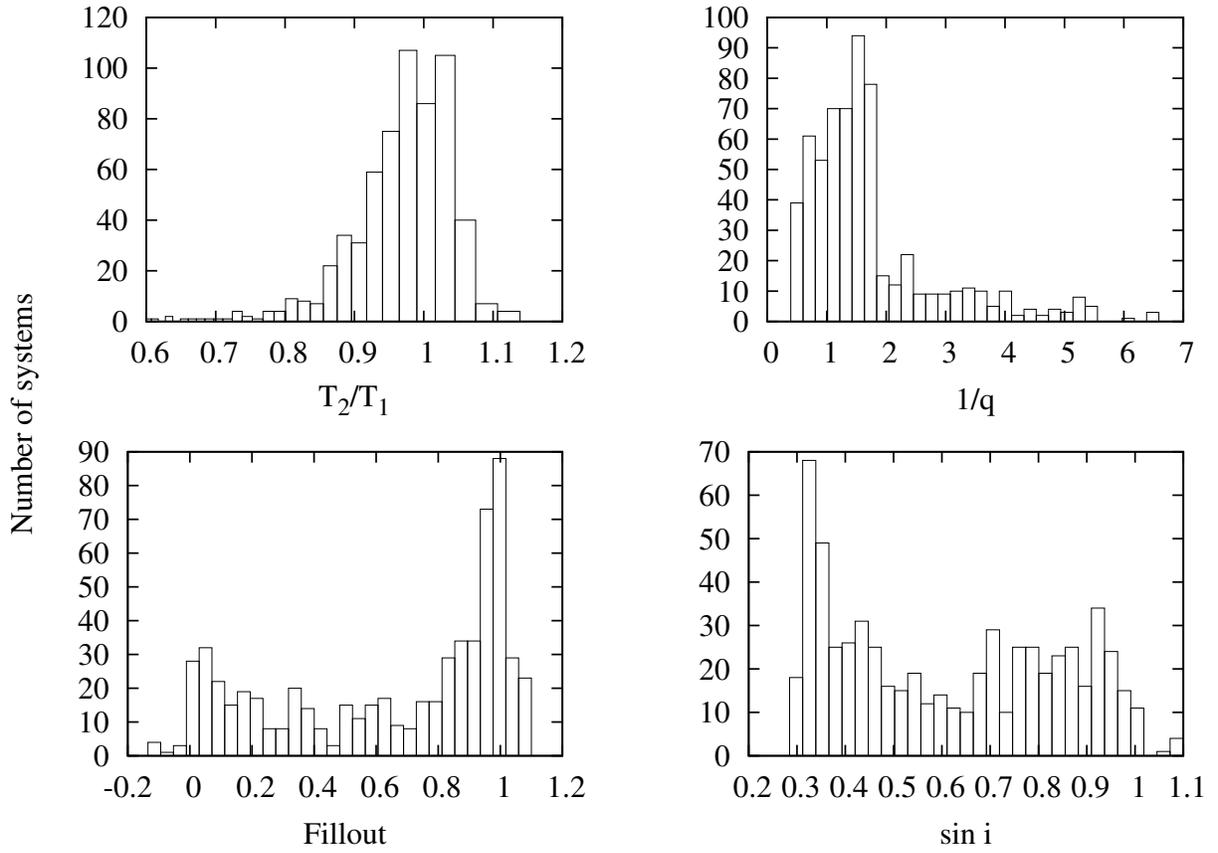} \\
\caption{The distributions of the 4 principal parameters for overcontact binaries and ellipsoidal variables. See text for discussion.} \label{fig_kepler_oc}
\end{center}
\end{figure}

\begin{figure}
\begin{center}
\includegraphics[height=\textwidth,angle=-90]{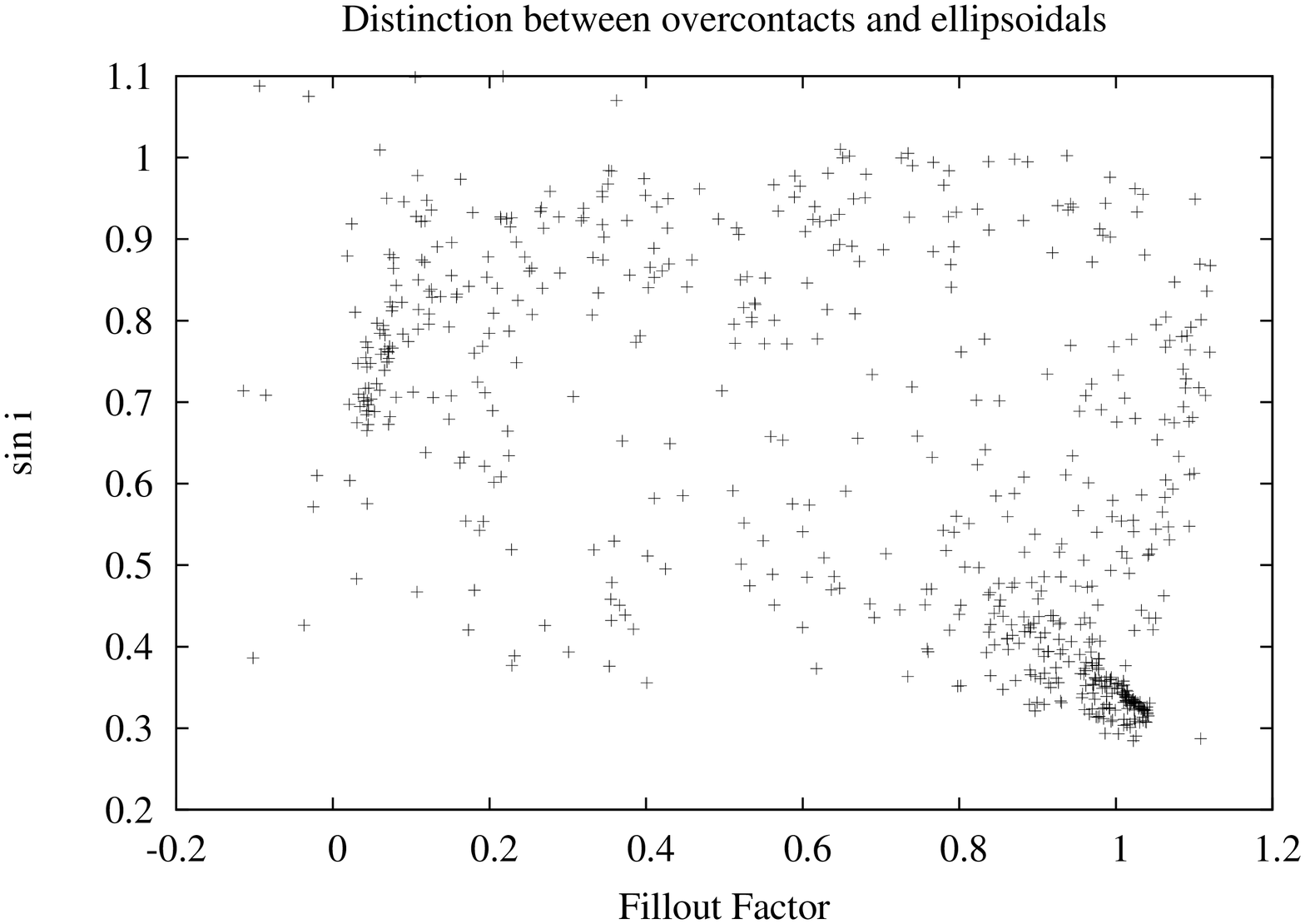} \\
\caption{The correlation between the fillout factor $F$ and $\sin i$. The area at bottom right corresponds to ellipsoidal variables.} \label{fig_correlation}
\end{center}
\end{figure}

Fig.~\ref{fig_kepler_oc} depicts the results for overcontact binaries. Most observed trends may be readily understood and interpreted.
\begin{description}
\item[$T_2/T_1$.] The temperature ratio peaks at $\sim$1, implying that most overcontact binaries are in thermal contact. A somewhat faster drop-off on a higher end is a consequence of our phasing convention that places the deeper eclipse at phase 0.
\item[$q$.] The mass ratio (depicted is the inverse $1/q$) peaks at values around unity. Algorithmically the code uses $1/q$ in place of $q$ for numerical stability: since surface potential $\Omega$ depends explicitly on $q$, as $q \to 0$, $\Omega(L_1) \to \Omega(L_2)$, causing singular solutions. This problem is solved easily by inverting the roles of stars and applying a 0.5 phase shift. The distribution gradually tails off to $q \sim 0.15$, values that are observed for extreme mass ratio overcontacts.
\item[$F$.] The fillout factor is expected to be roughly uniform, with a slight peak just below 1 because overcontacts are most readily detected in shallow contact. The strong peak at $F \sim 1.0$ can thus be surprising at first. However, closer inspection reveals that the peak consists of two distinct distributions: systems with inclinations across the whole dynamical range, corresponding to overcontacts and close-to-contact binaries, and those with low inclinations, corresponding to ellipsoidal variables. Fig.~\ref{fig_correlation} depicts the correlation between fillout factor and inclination that demonstrates this.
\item[$\sin i$.] Overcontacts can be detected even at low inclinations since their cross-sections change continuously with orbital phase due to surface distortion. We thus expect the distribution in $\sin i$ to be roughly uniform, and to tail off at low inclinations where ellipsoidal effect is diminished. The peak at low inclinations corresponds to ellipsoidal variables.
\end{description}

\begin{figure}
\begin{center}
\includegraphics[height=\textwidth,angle=-90]{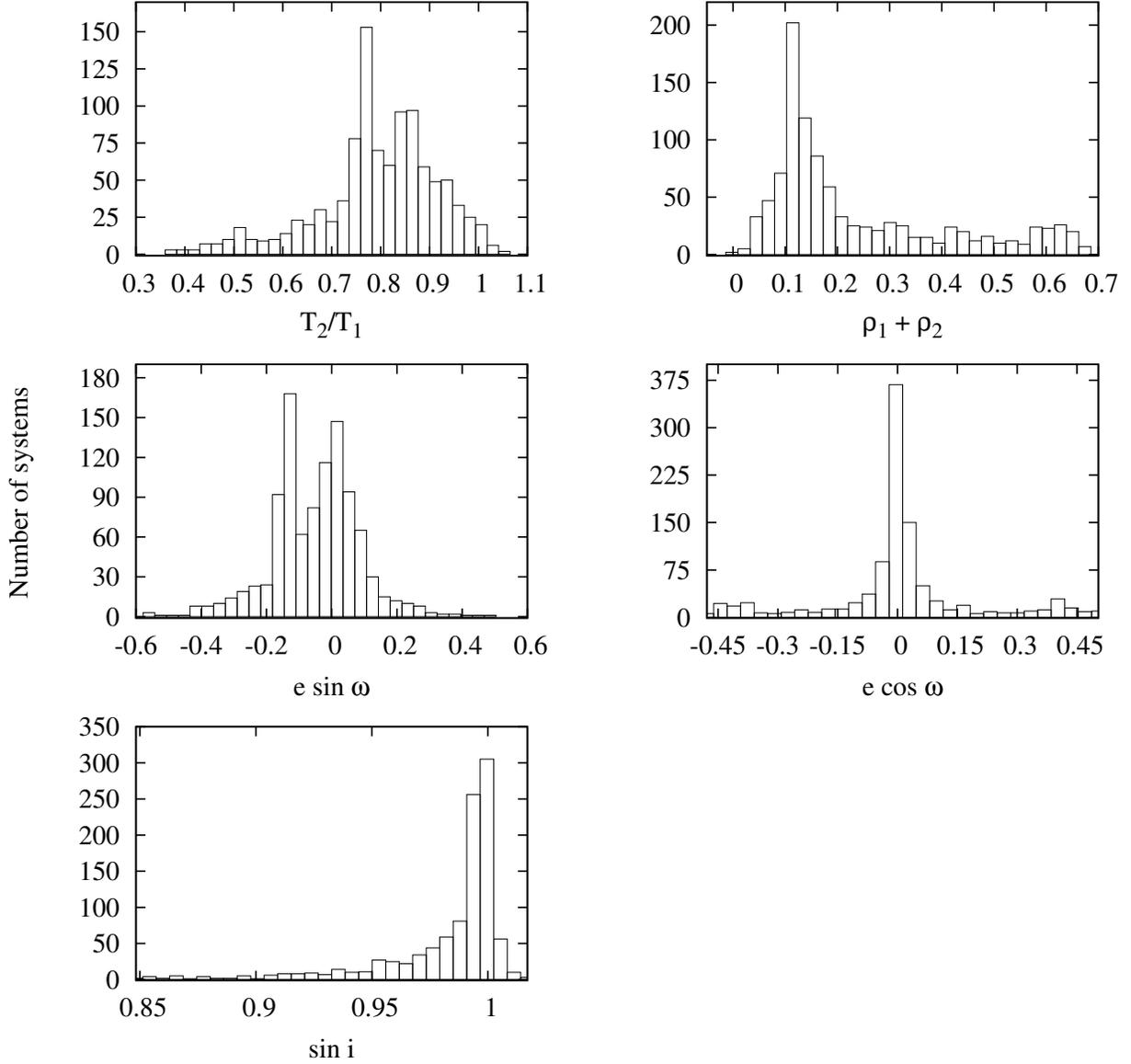} \\
\caption{The distributions of the 5 principal parameters for detached/semi-detached binaries. See text for discussion.} \label{fig_kepler_d}
\end{center}
\end{figure}

Parameter distributions for detached and semi-detached binaries are equally instructive. Fig.~\ref{fig_kepler_d} depicts the results.

\begin{description}
\item[$T_2/T_1$.] Temperature ratio peaks at values lower than 1.0, which would be 
expected. This is a consequence of systematics discussed in \S\ref{d_validation}: $T_2/T_1$ is a poor proxy to the surface brightness ratio. The inclusion of semi-detached binaries, which typically have mass ratios of $\sim 0.8$ or less, skews the distribution and might be responsible for the peak at $T_2/T_1 \sim 0.77$. \Kepler light curves often exhibit other types of variability that deteriorate the performance of the network, in particular with respect to $T_2/T_1$.
\item[$\rho_1+\rho_2$.] The sum of fractional radii reflects the selection effect of using only Q1 data for modeling: the longest period can be $\sim 20$-d, so we will be more susceptible to detecting close binaries than wide. As more quarters are included in our analysis, we can anticipate that the distribution will flatten out.
\item[$e$ {\rm and} $\omega$.] The orthogonalized components of eccentricity are distributed around 0, which corresponds to circular orbits, and disperse to the extent typical of eccentric binaries \citep{prsa2008}.
\item[$\sin i$.] The peak at $1.0$ and a fast drop-off are observed in the sample, in accordance with the geometrical requirements for systems to exhibit eclipses.
\end{description}

Although preliminary, these results should have statistical validity. As more data from \Kepler become available, these parameters will be refined and the light curves submitted to the physical modeling codes.

\section{Conclusion}

Hipparcos observed 917 EBs in a sample of 118,218 stars, which corresponds to a 0.8\% occurrence rate. Based on our conservative estimates, we are seeing a $\sim$1.2\% occurrence rate. This 50\% increase in detection attests to {\sl Kepler}'s photometric superiority. Even this rate might be slightly underestimated because of the dominant selection effect: short time-scale. While we have not been able to include long-period EBs in this catalog, such systems have already been noted in Q2 and Q3 data and will be published in a follow-up paper. Since \Kepler does not change its field, it is our aim to bring the catalog to fruition in steps, as more and more data become available. This paper, the first in a series, presents an extensive list of EBs and preliminary estimates of their principal parameters. It is our hope that the catalog will serve the eclipsing binary community as a bridge between the now public Q0/Q1 data and in-depth scientific modeling.

\acknowledgements

\Kepler was selected as the 10th mission of the Discovery Program. Funding for this mission is provided by NASA, Science Mission Directorate. We thank the entire \Kepler Mission team, and the data calibration engineers in particular, for making this possible.

This work is funded in part by the NASA/Caltech subcontract \#2-1085696 (PI Pr\v 
sa) and NSF RUI \#AST-05-07542. Doyle and Slawson are supported by the \Kepler {\it Mission} Participating Scientist Program, NASA grant NNX08AR15G awarded to LRD. Welsh acknowledges support from the \Kepler Participating Scientists Program via NASA grant NNX08AR14G.

We thank the following undergraduate and graduate students at San Diego State
University and Villanova University who assisted in ephemeris determination
and validation: Atul Belur, Tara Fetherolf, Trevor Ames Gregg, Steve 
Shipler, Nicole Zimmerman and Tim McClain. We also thank Ron Angione, John Sievers, and Paul Etzel for their assistance and advice.

{\it Facilities:} \facility{The \Kepler Mission}

\bibliographystyle{apj}
\bibliography{paper.rev2}

\end{document}